\documentclass[preprint,aps,12pt,showpacs,nofootinbib,tightenlines]{revtex4}
\usepackage{amsmath}
\usepackage{amssymb}
\usepackage{epsfig}
\usepackage{graphicx}
\begin{document}
\preprint{NJNU-TH-2004-11}
\def \epjc{  Eur. Phys. J. C }
\def \jpg{  J. Phys. G }
\def \mpla{ Mod.Phys.Lett. A }
\def \npb{  Nucl. Phys. B }
\def \plb{  Phys. Lett. B }
\def \prd{  Phys. Rev. D }
\def \prl{  Phys. Rev. Lett.  }
\def \pr{   Phys. Rep. }
\def \rmp{  Rev. Mod. Phys. }

\newcommand{\bequ}{\begin{equation}}
\newcommand{\eequ}{\end{equation}}

\newcommand{\beq}{\begin{eqnarray}}
\newcommand{\eeq}{\end{eqnarray}}

\newcommand{\etap}{\eta'}
\newcommand{\tab}[1]{Table \ref{#1}}
\newcommand{\fig}[1]{Fig.\ref{#1}}
\newcommand{\real}{{\rm Re}\,}
\newcommand{\im}{{\rm Im}\,}
\newcommand{\non}{\nonumber\\ }


\title{ Charmless $B \to PP $ decays and the new physics effects
in the minimal supergravity model  }
\author{ Zhenjun Xiao}
\email{xiaozhenjun@pine.njnu.edu.cn}
\affiliation{Department of Physics, Nanjing Normal University, Nanjing,
Jiangsu 210097, P.R.China}
\affiliation{ CCAST(World Laboratory), P.O.Box 8730, Beijing 100080, China}
\author{Wenjuan Zou}
\affiliation{Department of Physics, Nanjing Normal University,
Nanjing, Jiangsu 210097, P.R.China}
\date{\today}
\begin{abstract}
By employing the QCD factorization approach, we calculate the new
physics contributions to the branching radios of the
two-body charmless $ B \to PP$ decays in the framework of the minimal
supergravity (mSUGRA) model. Within the considered parameter space, we
find that (a) the supersymmetric (SUSY) corrections to the
Wilson coefficients $C_k$ ($k=3-6$)
are very small and can be neglected safely, but the leading order SUSY
contributions  to  $C_{7\gamma}(M_W)$ and $C_{8g}(M_W)$ can be rather large
and even change the sign of the corresponding coefficients in the standard
model;
(b) the possible SUSY contributions to those penguin-dominated decays
in mSUGRA model can be as large as $30-50\%$;
(c) for the well measured $B \to K \pi$ decays,
the significant SUSY contributions play an important rule to improve the
consistency of the theoretical predictions with the data;
(d) for $B \to K \eta'$ decays, the theoretical predictions of the corresponding
branching ratios become consistent with the data within one standard deviation
after the inclusion of the large SUSY contributions in the mSUGRA model.
\end{abstract}


\maketitle

\section{Introduction}

As is well known,  the precision measurements of the B meson system can
provide an insight into very high energy scales via the indirect loop
effects of the new physics beyond the standard model (SM)
\cite{slac504,hurth03}.
Although currently available data agree well with the SM predictions, we
generally believe that the B-factories can at least detect the first
signals of new physics if it is there.

Among the $B \to P P$ (P stands for the pseudo-scalar light
mesons) decay channels considered in this paper, twelve of them
have been measured with good accuracy. And the data indeed show
some deviations from the SM expectations:
\begin{itemize}

\item
The $K \etap$ puzzle,  the observed $B \to K\etap$
branching ratios \cite{cleo03,babar03,belle03}
are much larger than the corresponding SM predictions,
appeared several years ago, and there is
still no convincing theoretical interpretation for this puzzle
after intensive studies in the framework of SM \cite{as97}
and the new physics models\cite{etapnp}.

\item The $K \pi$ puzzle comes from the ratios $R_c$ and $R_n$ for
the four well measured $B \to K \pi$ decay rates as defined in
Ref.~\cite{gr03}. The SM prediction is $R_c = R_n$ by neglecting
the small exchange- and annihilation-type amplitudes \cite{gr03},
while the present data \cite{hfag04} yields \beq R_c^{exp}&=&
\frac{2 \Gamma(B^+ \to K^+ \pi^0)}{\Gamma(B^+ \to K^0 \pi^+)}
= 1.15 \pm 0.13, \\
R_n^{exp} &=&\frac{ \Gamma(B_d^0 \to K^+ \pi^-)}{2\Gamma(B_d^0 \to K^0 \pi^0)}
= 0.78 \pm 0.10
\eeq
A discrepancy of $2.8 \sigma$ exist here.

\item
For $B \to \pi^0 \pi^0$ decay, the measured branching ratio
$Br(B \to \pi^0 \pi^0)
=(1.9 \pm 0.5)\times 10^{-6}$ \cite{babar03,belle03,hfag04} is about five
times larger than the SM prediction.

\end{itemize}
Although not convincing, these discrepancies together with the so-called
$\phi K_s$ anomaly \cite{phiks} may be the first hints
of new physics beyond the SM in B experiments \cite{london04,sil04}.

Up to now, the possible new physics contributions to rare
B meson decays have been studied extensively, for example,
in the Technicolor models\cite{tc2},
the two-Higgs-doublet models\cite{etapm3,2hdm1}
and the supersymmetric models \cite{hmdx98,tyy96,tyy97,mssm}.
Among the various new physics models, the supersymmetric models
are indeed the most frequently studied models in searching for new physics in
B meson system.
The minimal supersymmetric standard model (MSSM)\cite{as99} is the
general and most economical low-energy supersymmetric extension of the
SM. But it is hardly to make definite predictions
for the physical observables in B meson decays since there are more than
one hundred free parameters appeared in the MSSM.
In order to find the possible signals or hints of new physics beyond the
SM from the data, various scenarios of the MSSM are proposed by imposing
different constraints on it \cite{as99}.
The minimal supergravity (mSUGRA) model \cite{msugra} seems to be a very simple
constrained MSSM  model, since it have only five free parameters $ \tan\beta,
m_{\frac{1}{2}}, m_{0},A_{0}$, and $sign(\mu)$ at the high energy scale.

The previous works in the framework of mSUGRA model focused
on the semileptonic, leptonic and radiative rare B decays.
In Refs.\cite{hmdx98,tyy96,tyy97,huang03}, for example, the authors studied the
rare decays $B\to X_s \gamma$, $B \to X_s l\bar{l}$, $B \to l^+ l^-$ and
the  $B^{0}-\bar{B^{0} }$ mixing
in the mSUGRA model, and found some constraints on the parameter space of this
model.

For $B \to P P$ decays, they have been studied in the SM
\cite{ali9804,chen99,bbns99,mgmc01,mm03,du02,pqcd},
the Technicolor models\cite{tc2} and
the two-Higgs-doublet models\cite{etapm3}.
In Ref.\cite{sa04}, Mishima and Sanda calculated the supersymmetric
effects on $B\to \phi k$ decays in the PQCD approach\cite{pqcd}
and predicted the values of CP asymmetries with the inclusion of the
supersymmetric contribution.
In this paper, we calculate the supersymmetric contributions to the branching radios
of the twenty one  $ B \to PP$  decay modes in the mSUGRA model
by employing the QCD factorization approach (QCD FA) \cite{bbns99,mgmc01,mm03}.
The contributions from chirally enhanced power
corrections and weak annihilations are also taken into account.
We find that the branching ratios of some decay modes can be enhanced
significantly, and these new contributions can help us to
give a new physics interpretation for the so-called $`` K \eta'  "$ puzzle.

This paper is organized as follows. In section II, we give a brief review
for the minimal supergravity model.
In section III, we calculate the new penguin
diagrams induced by new particles and extract out  the new physics parts
of the Wilson coefficients in the mSUGRA model.
The calculation of $B\to PP$ decays in
QCD factorization approach  is also discussed in this section. In
section IV,  we present the numerical results of the branching ratios for the
twenty one  $B\to PP$ decay modes in the SM and the mSUGRA model,
and make phenomenological analysis for those well measured decay modes.
The final section is the summary.

\section{Outline of the mSUGRA model}   \label{sec:msugra}

In the MSSM, the most general superpotential compatible with gauge invariance,
renormalizability and R-parity conserving is written as \cite{as99}:
\beq
{\cal W}=\varepsilon_{\alpha\beta}\left [f_{Uij}Q_{i}^{\alpha}H_{2}^{\beta}U_{j}
                           +f_{Dij}H_{1}^{\alpha}Q_{i}^{\beta}D_{j}
                           +f_{Eij}H_{1}^{\alpha}L_{i}^{\beta}E_{j}
                           -\mu H_{1}^{\alpha}H_{2}^{\beta} \right ]
\eeq
where $f_{D}$, $f_{U}$ and $f_{E}$ are Yukawa coupling constants
for down-type, up-type quarks, and leptons, respectively. The
suffixes $\alpha,\beta=1,2$ are SU(2) indices and i,j=1,2,3 are
generation indices, $\varepsilon_{\alpha\beta}$ is the
antisymmetric tensor with $ \varepsilon_{12}=1$. In addition
to the SUSY invariant terms, a set of terms which explicitly but softly break
SUSY should be added to the supersymmetric Lagrangian. A general
form of the soft SUSY-breaking terms is given as \cite{as99}:
\beq
-{\cal L}_{soft}&=&
    \left (m^{2}_{Q}\right )_{ij}\tilde{q}^{+}_{Li}\tilde{q}_{Lj}
   +\left(m^{2}_U\right )_{ij}\tilde{u}^{*}_{Ri}\tilde{u}_{Rj}
   +\left(m^{2}_D\right )_{ij}\tilde{d}^{*}_{Ri}\tilde{d}_{Rj}
   +\left(m^{2}_L\right )_{ij}\tilde{l}^{+}_{Li}\tilde{l}_{Lj}\non
&\ \ &
   +\left(m^{2}_E\right )_{ij}\tilde{e}^{*}_{Ri}\tilde{e}_{Rj}
   +\Delta^{2}_{1}h_{1}^{+}h_{1}+\Delta^{2}_{2}h_{2}^{+}h_{2} \non
&\ \ &
   +\varepsilon_{\alpha\beta}
   \left [A_{Uij}\tilde{q}^{\alpha}_{Li}h^{\beta}_{2}\tilde{u}^{*}_{Rj}
   +A_{Dij}h^{\alpha}_{1}\tilde{q}^{\beta}_{Li}\tilde{d}^{*}_{Rj}
   +A_{Eij}h^{\alpha}_{1}\tilde{l}^{\beta}_{Li}\tilde{e}^{*}_{Rj}
   +B\mu h^{\alpha}_{1}h^{\beta}_{2}\right ]\non
&\ \ &
   +\frac{1}{2}m_{\tilde{B}}\tilde{B}\tilde{B}
   +\frac{1}{2}m_{\tilde{W}}\tilde{W}\tilde{W}
   +\frac{1}{2}m_{\tilde{G}}\tilde{G}\tilde{G} + H.C.
\label{eq:lsoft}
\eeq
where $\tilde{q}_{Li}$, $\tilde{u}^{*}_{Ri}$, $\tilde{d}^{*}_{Ri}$,
$\tilde{l}_{Li}$, $\tilde{e}^{*}_{Ri}$, and $h_1$ and $h_2$ are
scalar components of chiral superfields $Q_i$, $U_i$, $ D_{i}$,
$L_{i}$, $E_{i}$, $H_1$, and $H_2$ respectively, and $\tilde{B}$,
$\tilde{W}$, and $\tilde{G}$ are $ U(1)_Y$, $SU(2)_L$, and $
SU(3)_C $ gauge fermions. And the terms appeared in Eq.(\ref{eq:lsoft})
are the mass terms for the scalar fermions, mass
and bilinear terms for the Higgs bosons, trilinear coupling terms
between sfermions and Higgs bosons, and mass terms for the
gluinos, Winos and binos, respectively.

In the mSUGRA model, a set of assumptions are added to the MSSM. One underlying
assumpsion is that SUSY-breaking occurs in a hidden sector which
communicates with the visible sector only through gravitational
interactions.  The free parameters in the MSSM are assumed to obey a set of
boundary conditions at the Plank or GUT scale:
\beq
 \alpha_{1}&=&\alpha_{2}=\alpha_{3}=\alpha_{X}, \non
(m^{2}_{Q})_{ij}&=& (m^{2}_{U})_{ij}=(m^{2}_{D})_{ij}=(m^{2}_{L})_{ij}
=(m^{2}_{E})_{ij}=(m^{2}_{0})\delta_{ij}, \non
\Delta^{2}_{1}&=&\Delta^{2}_{2}=m^{2}_{0}, \non
A_{Uij}&=&f_{Uij}A_{0},\ \ A_{Dij}=f_{Dij}A_{0}, \ \ A_{Eij}=f_{Eij}A_{0}, \non
m_{\tilde{B}}&=& m_{\tilde{W}}=m_{\tilde{G}}=m_{\frac{1}{2}}
\eeq
where $\alpha_{i}=g^2_i/(4\pi)$, and  $g_{i}$ (i=1,2,3) denotes
the coupling constant of the $U(1)_Y$, $SU(2)_L$, $SU(3)_C$
gauge group, respectively. The unification of them is verified according to the
experimental results from LEP1\cite{pdg02} and  can be  fixed
at the Grand Unification Scale $M_{GUT}\sim 2\times 10^{16}Gev$.
Besides the three parameters $ m_{\frac{1}{2}}$, $m_{0}$ and
$A_{0}$, the supersymmetric sector is described at GUT scale by
the bilinear coupling B and the supersymmetric Higgs(ino) mass
parameter $\mu$. However, one has to require the radiative
electroweak symmetry-breaking (EWSB) takes place at the low energy
scale. The effective potential of neutral Higgs fields at the
tree-level is given by (to be precise, one-loop corrections to the scalar
potential have  been included in the program we used later)
\beq
V_{Higgs}&=&m^{2}_{1}|h^{0}_{1}|^2+m^{2}_{2}|h^{0}_{2}|^2+m^{2}_{3}(h_{1}^0h_{2}^0+H.C)\nonumber\\
&  &+\frac{g^{2}_{1}+g^{2}_{2}}{8}(|h^{0}_{1}|^2-|h^{0}_{2}|^2)^{2}
\label{eq:vhiggs}
\eeq
where we have used the usual short-hand notation: $m^{2}_{1}=(\mu^{2}+\Delta^{2}_{1})$,
$m^{2}_{2}=(\mu^{2}+\Delta^{2}_{2})$, $m^{2}_{3}=B\mu$. The radiative EWSB condition is
\beq
\label{eq:vv}
\langle\frac{\partial V}{\partial
h^{0}_{1}}\rangle=\langle\frac{\partial V}{\partial
h^{0}_{2}}\rangle=0
\eeq
where the value $h^{0}_{1}$, $h^{0}_{2}$ denotes the vacuum
expectation values of the two neutral Higgs fields as $\langle
h^{0}_{1} \rangle=v\cos\beta$, $\langle h^{0}_{2} \rangle=v\sin\beta$
with $v=174 Gev$. From Eq.(\ref{eq:vv}), we can
determine the values of $\mu^{2}$ and $B\mu$:
\beq
\label{eq:bb}
\mu^{2}&=&\frac{1}{2}[\tan 2\beta(\Delta^{2}_{2}\tan\beta-\Delta^{2}_{1}\cot\beta)-M_{Z}^{2}]\nonumber\\
B\mu&=&\frac{1}{2}\sin
2\beta[\Delta^{2}_{1}+\Delta^{2}_{2}+2\mu^{2}]
\eeq
Through Eq.(\ref{eq:bb}) we can see the sign of $\mu$ is not determined.
Therefore only four continuous free parameters, and an unknown
sign is left in the mSUGRA model. They are:
\beq
\tan\beta, m_{\frac{1}{2}}, m_{0},A_{0},sign(\mu)
\eeq

In the mSUGRA model, all other parameters at the electroweak scale
are then determined through the five free parameters by the GUT
universality and the renormalization group equation (RGE) evolution.
In this paper, we calculate the SUSY and Higgs particle spectrum
through a Fortran code: SUSPECT version 2.1 \cite{ajg02}. The important
features of this code include
(a) the renormalization group evolution between low and high energy scales;
(b) consistent implementation of radiative electroweak symmetry breaking;
and (c) calculation of the physical particle masses with radiative corrections.
Using this code, we obtain the SUSY and Higgs particle masses, and the mixing angles
of squarks at the electroweak scale. From these Low-energy supersymmetric parameters,
the mixing matrices $\Gamma^{U}$,$\Gamma^{D}$ for the up-type and the down-type
squarks, the mixing matrices $U, V, N$ for charginos and neutralinos are determined.
The explicit expressions of the two $6\times 6$ mixing matrices
$\Gamma^{U}$ and $\Gamma^{D}$, two
$2\times 2$ matrices $U$ and $V$, and a $4 \times 4$ matrix $N$
can be found in Refs.\cite{ajg02,ctfj,pm96}.

\section{The basic theoretical framework for $ B\to PP$ }
\label{sec:tbtf}

In this section, we present the theoretical framework and the
relevant formulas for calculating the  exclusive nonleptonic
decays of the $ B^{\pm}$ and $ B^{0}$ mesons into two light
pseudoscalar mesons.

\subsection{Effective Hamiltonian and relevant Wilson coefficients
in SM }

In the SM, if we take into account only the operators
up to dimensions $6$, and assume $m_b\gg m_s$, the
effective Hamiltonian for the quark level three-body decay
$ b\to qq^{'}\bar{q^{'}}$ $(q\in\{d,s\}, q^{'}\in \{ u,d,s\})$
at the scale $\mu$ reads \cite{ajb98}
\beq
{\cal H}_{eff}&=&\frac{G_{F}}{\sqrt{2}}\left\{\sum_{i=1}^{2}C_{i}(\mu)
\left [V_{ub}V_{uq}^{*}O_{i}^{u}(\mu)+V_{cb}V_{cq}^{*}O_{i}^{c}(\mu)\right ] \right.\non
 &  & \left. -V_{tb}V_{tq}^{*}\sum_{j=3}^{10}C_{j}(\mu)O_{j}(\mu)
             -V_{tb}V_{tq}^{*}\left [C_{7\gamma}(\mu)O_{7\gamma}(\mu)+C_{8g}(\mu)O_{8g}(\mu)
             \right ] \right \}
\label{eq:eff} \eeq where $V_{pb}V_{pq}^{*}$ is the products of
elements of the Cabbibo-Kabayashi-Maskawa quark mixing
matrix\cite{ckm}. And the current-current ($O_{1,2}$), QCD penguin
($O_{3,4,5,6}$), electroweak penguin ($O_{7,8,9,10}$),
electromagnetic and chromomagnetic dipole operators ($O_{7\gamma}$
and $O_{8g}$) can be written as \cite{gam96}
\begin{align}
O_{1}^{u}&=(\bar{q}u)_{V-A}(\bar{u}b)_{V-A},&
O_{2}^{u}&=(\bar{q}_{\alpha}u_{\beta})_{V-A}(\bar{u_{\beta}}b_{\alpha})_{V-A},
\nonumber\\
O_{1}^{c}&=(\bar{q}c)_{V-A}(\bar{c}b)_{V-A},&
O_{2}^{c}&=(\bar{q}_{\alpha}c_{\beta})_{V-A}(\bar{c_{\beta}}b_{\alpha})_{V-A},
\nonumber\\
O_{3}&=(\bar{q}b)_{V-A}\sum_{q^{'}}(\bar{q}^{'}q^{'})_{V-A},&
O_{4}&=(\bar{q}_{\alpha}b_{\beta})_{V-A}\sum_{q^{'}}(\bar{q_{\beta}}^{'}q^{'}_{\alpha})_{V-A},
\nonumber\\
O_{5}&=(\bar{q}b)_{V-A}\sum_{q^{'}}(\bar{q}^{'}q^{'})_{V+A},&
O_{6}&=(\bar{q}_{\alpha}b_{\beta})_{V-A}\sum_{q^{'}}(\bar{q_{\beta}}^{'}q^{'}_{\alpha})_{V+A},
\nonumber\\
O_{7}&=\frac{3}{2}(\bar{q}b)_{V-A}\sum_{q^{'}}e_{q^{'}}(\bar{q}^{'}q^{'})_{V+A},&
O_{8}&=\frac{3}{2}(\bar{q}_{\alpha}b_{\beta})_{V-A}\sum_{q^{'}}e_{q^{'}}(\bar{q_{\beta}}^{'}q^{'}_{\alpha})_{V+A},
\nonumber\\
O_{9}&=\frac{3}{2}(\bar{q}b)_{V-A}\sum_{q^{'}}e_{q^{'}}(\bar{q}^{'}q^{'})_{V-A},&
O_{10}&=\frac{3}{2}(\bar{q}_{\alpha}b_{\beta})_{V-A}\sum_{q^{'}}e_{q^{'}}(\bar{q_{\beta}}^{'}q^{'}_{\alpha})_{V-A},
\nonumber\\
O_{7\gamma}&=\frac{e}{8\pi^{2}}m_{b}\bar{q}_{\alpha}\sigma^{\mu\nu}(1+\gamma_{5})b_{\alpha}F_{\mu\nu},&
O_{8g}&=\frac{g}{8\pi^{2}}m_{b}\bar{q}_{\alpha}\sigma^{\mu\nu}(1
+\gamma_{5})T^a_{\alpha\beta}b_{\beta}G^{a}_{\mu\nu},
\end{align}
where $T^{a}(a=1,...,8)$ stands for $SU(3)_{c}$ generators,
$\alpha$ and $\beta$ are the $SU(3)_C$ color indices,
and $V\pm A\equiv \gamma_{\mu}(1\pm\gamma_{5})$ by definition.
The sum over $q'$ runs over the quark fields that are active at the scale
$\mu={\cal O}(m_b)$, i.e., $q'\in \left \{ u,d,s,c,b\right \} $.

To calculate the non-leptonic B meson decays at next-to-leading order in
$\alpha_{s}$ and to leading power in $\Lambda_{QCD}/m_{b}$, we
should determinate the Wilson coefficient $C_{i}(M_{W})$ through
matching of the full theory onto the five-quark low energy
effective theory where the $W^\pm$ gauge boson, top quark and all SUSY particles
heavier than $M_{W}$ are integrated out, and run the
Wilson coefficients down to the low energy scale $\mu\sim O(m_b)$
by using the QCD renormalization group equations. In table \ref{tab:cimb},
we simply present the numerical results of the LO and NLO Wilson coefficient
in the NDR scheme in different scales. More detailed analytical expressions can be
found for example in Refs.\cite{ajb98,gam96}.

\begin{table}[htb]
\doublerulesep 1.5pt
\caption{In the NDR scheme, the values of LO
and NLO Wilson coefficients $C_{i}(\mu)$ for $\mu=m_b/2, m_b, 2m_b$.
Input parameters being used are $\Lambda^{(5)}_{\overline{ms}}=0.225Gev,
sin^{2}\theta_{W}=0.23, m_b=4.62Gev, m_t=175Gev, M_W= 80.4Gev$, and
$\alpha_{em}=1/128$.}
\label{tab:cimb}
\begin{center}
\begin{tabular} {c|rr|rr|rr} \hline  \hline
\multicolumn{1}{c|}{ \  \ } & \multicolumn{2}{|c|}{$\mu=m_{b}/2$} &
\multicolumn{2}{|c|}{$\mu=m_{b}$} & \multicolumn{2}{|c}{$\mu=2m_{b}$}
 \\ \cline{2-7}
\  \ & LO & NLO & LO & NLO & LO & NLO  \\ \hline \hline
$C_1$  &1.179&1.134&1.115&1.080&1.072&1.043 \\
$C_2$  &$-0.370$&$-0.280$&$-0.255$&$-0.180$&$-0.171$&$-0.104$   \\
$C_3$  &0.019& 0.020&0.012&0.013 &0.008&0.008      \\
$C_4$  &$-0.037$&$-0.048$ &$-0.027$&$-0.034$&$-0.018$&$-0.023$   \\
$C_5$  & 0.010&0.012&0.008&0.010 &0.006&0.007    \\
$C_6$  &$-0.050$&$-0.062$&$-0.033$&$-0.040$&$-0.021$&$-0.026$  \\
$C_{7}/\alpha_{em}$ &0.018&$-0.008$&0.028 &0.007&0.046&0.030 \\
$C_{8}/\alpha_{em}$ & 0.055&0.055&0.035  &0.035&0.023& 0.023 \\
$C_{9}/\alpha_{em}$ &$-1.398$&$-1.420$&$-1.318$  & $-1.337$&$-1.255$&$-1.270$\\
$C_{10}/\alpha_{em}$ &0.415&0.395&0.286&0.273&0.191&0.183\\
$C_{7\gamma}$& $-0.360 $&$-0.334$& $-0.316$&$-0.307 $ & $-0.281$&$-0.282$ \\
$C_{8g}$     &$-0.167$& \   \ &$-0.150$& \   \ & $-0.136$&\   \ \\
\hline \hline
\end{tabular}
\end{center}
\end{table}

\subsection{Wilson coefficients in the mSUGRA model  }

In the mSUGRA model, the new physics contributions to the rare decays will
manifest themselves through two channels. One is the new
contributions to the Wilson coefficients of the same operators
involved in the SM calculation, the other is to the Wilson
coefficients of the new operators  such as operators with opposite
chiralities. In the SM, the latter is absent because they are
suppressed by the ratio $m_{s}/m_{b}$. In the mSUGRA model, they
can also be neglected, as shown in Ref.\cite{ctf02}. Therefore we
here use the same operator base as in the SM.

It is well known that there is no SUSY contributions to the Wilson
coefficients at the tree level. There are five kinds of
contributions to the quark level decay process $ b\to qq^{'}\bar{q^{'}}$
at one-loop level, depending on specific particles propagated in
the loops:
\begin{itemize}
\item
the gauge boson $W^{\pm}$ and up-type quarks $u, c, t$, which leads to the
contributions in the SM;

\item
the charged Higgs boson $H^{\pm}$ and up-type quarks $u,c,t$;

\item
the charginos $\tilde{\chi}^{\pm}_{1,2}$ and the scalar up-type quarks
$\tilde{u}, \tilde{c},\tilde{t}$;

\item
the neutralinos $\tilde{\chi}^{0}_{1,2,3,4}$ and the down-type squarks
$\tilde{d}, \tilde{s},\tilde{b}$;

\item the gauginos $\tilde{g}$ and the down-type squarks
$\tilde{d}, \tilde{s},\tilde{b}$.

\end{itemize}

The new physics contributions from those superparticle loops may induce
too large flavor changing neutral currents (FCNCs). To escape from the
so-called SUSY flavor problem, degeneracy of masses of squarks and
sleptons among different generations has been assumed in the
minimal SUGRA model.

\begin{figure}[]
\vspace{-1cm}
\centerline{\mbox{\epsfxsize=18cm\epsffile{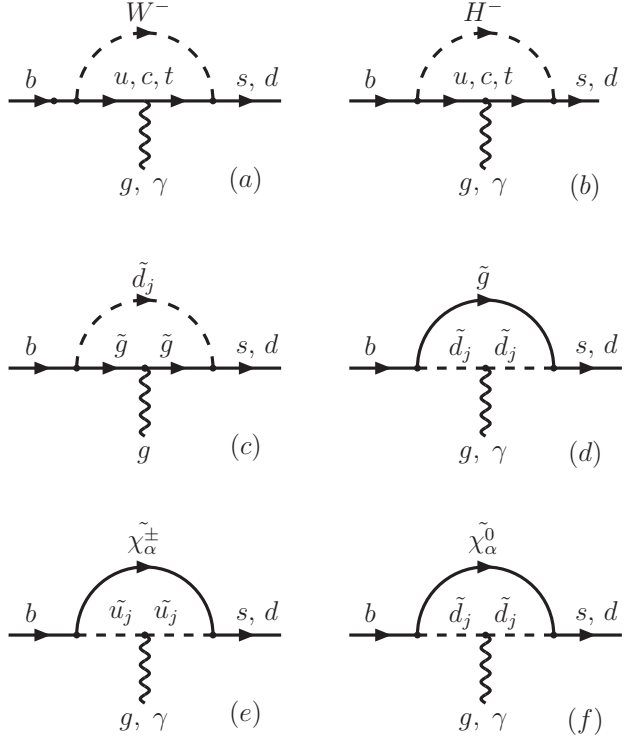}}}
\vspace{-12cm}
\caption{Five kinds of loop contributions to $ b \to
qg, q\gamma $: (a) SM contribution. (b) charged Higgs contribution
; (c-d) gluino contribution; (e) chargino
contribution; (f) neutralino contribution.} \label{fig:feynman}
\vspace{20pt}
\end{figure}

In order to determine the new physics contributions
to Wilson coefficients $C_{i} (i=3,4,5,6)$, $C_{7\gamma}$, and $C_{8g}$
(we ignore the new physics contributions to $C_{i} (i=7,8,9,10)$ because they are
suppressed by a factor of $\alpha_{em}/\alpha_{s}$ ) at the $M_w$ scale,
we need to calculate the Feynman diagrams appeared in Fig.\ref{fig:feynman}.
First, by employing conservation of the gluonic current,
we can define the effective vertex of the $b\to q g$ penguin processes
as in Ref.\cite{swi98}:
\beq
\Gamma^{a}_{\mu}(q^{2})&=&\frac{ig_{s}}{4\pi^{2}}\bar{u}_{q}(p_{q})T^{a}V_{\mu}(q^{2})u_{b}(p_{b})
\eeq
with
\beq
V_{\mu}(q^{2})&=&(q^{2}g_{\mu\nu}-q_{\mu}q_{\nu})\gamma^{\nu}
\left [F_{1L}(q^2)P_{L}+F_{1R}(q^{2})P_{R}\right ]\non
& \  \ & +  i\sigma_{\mu\nu}q^{\nu}\left [F_{2L}(q^2)P_{L}+F_{2R}(q^{2})P_{R}\right ]
\eeq
where $F_{1}(q^2)$ and $F_{2}(q^2)$ are the electric and magnetic
 form factors, $q=p_{b}-p_{q}$ is the gluon momentum, and
$P_{L(R)}\equiv(1\pm \gamma_{5})/2$ are the chirality projection
operators.

By calculating the Feynman diagrams as shown in
Figs.\ref{fig:feynman}(b)-(f), we find (in the naive dimensional
regularization (NDR) scheme,) the new physics parts of the Wilson
coefficients at the scale $M_W$
\beq \label{eq:ck}
C_k^{NP}(M_{W})&=&-\frac{\alpha_{s}(M_{W})}{24\pi} \left
[\frac{G_F}{\sqrt{2}}\lambda_{t}\right ]^{-1}A_{k}F_{1L}(0) \\ \label{eq:c8g}
C_{8g}^{NP}(M_{W})&=&-\frac{F_{2R}(0)}{2}
\left [\frac{G_F}{\sqrt{2}}\lambda_{t}m_{b} \right ]^{-1}
\eeq
where $A_{k}\equiv\{ -1,3,-1,3 \}$ for $k=\{3,4,5,6\}$, and
$\lambda_{t}=V_{tq}^{*}V_{tb}$. In addition, since  $q^{2} \ll
m^{2}_{\tilde{q}}$ where $m_{\tilde{q}}$ is the mass of the heavy
scalar fermions, we set $q^2=0$ for the form factors $F_{1,2}$
as an approximation
\footnote{Eqs.(\ref{eq:ck}) and (\ref{eq:c8g}) differ from those
appeared in Ref.\cite{whi99}, but our final analytic expressions
for  $ C_{8g}^{NP}(M_{W})$ are the same as that in Ref.\cite{tyy97} except for
the definition for Wilson coefficients.}.
The explicit expressions  of the form factors $F_{1L}(0)$ and $F_{2R}(0)$
induced by supersymmetric particles are the following
\beq
\label{eq:f1lh}
F^{H^{-}}_{1L}(0) &=& -\frac
{G_{F}}{\sqrt{2}} \lambda_{t} x_{th} \cot^2{\beta} f_{5}(x_{th}),
\\
\label{eq:f2rh}
F^{H^{-}}_{2R}(0) &=&  \frac {G_{F}}{\sqrt{2}} \lambda_{t} m_{b} x_{th}
\left [\cot^{2}\beta f_{1}(x_{th})+f_{3}(x_{th}) \right ],
\\
F^{\tilde{g}}_{1L}(0) &=& -\frac
{g^{2}_{s}}{4m^{2}_{\tilde{g}}}\sum_{I=1}^{6}
\left (\Gamma^{d+}_{GL} \right )^{I}_{j}
\left (\Gamma^{d}_{GL}\right )^{3}_{I}f_6
\left (x_{\tilde{d_{I}}\tilde{G}} \right ),
\\
F^{\tilde{g}}_{2R}(0) &=&\sqrt{2}G_{F}\frac{g^{2}_{s}}{g^{2}}
\sum_{I=1}^{6}x_{W\tilde{d}_{I}}\left (\Gamma^{d+}_{GL}\right )^{I}_{j}
\left \{ \left (\Gamma^{d}_{GL} \right )^{3}_{I} m_{b}
\left [ 3f_{1} \left (x_{\tilde{G}\tilde{d_{I}}}\right )
+ \frac{1}{3}f_{2}\left (x_{\tilde{G}\tilde{d_{I}}}\right ) \right ] \right.
 \nonumber \\
 &  & \left.
 +\left (\Gamma^{d}_{GL} \right )^{3}_{I}m_{\tilde{G}}
 \left [ 3 f_{3} (x_{\tilde{G}\tilde{d_{I}}})+ \frac{1}{3}
 f_{4}(x_{\tilde{G}\tilde{d_{I}}}) \right ] \right \},
\\
F^{\chi^{-}}_{1L}(0) &=&-\frac{G_F}{\sqrt{2}}\sum_{\alpha=1}^{2}
\sum_{I=1}^{6}x_{W\tilde{u}_{I}}
\left (\Gamma^{d+}_{CL} \right )^{I}_{\alpha j}
\left (\Gamma^{d}_{CL} \right )^{\alpha 3}_{I}
f_{7}\left (x_{\tilde{x}_{\alpha}^{-}\tilde{u}_{I}}\right ), \\
F^{\chi^{-}}_{2R}(0) &=&-\sqrt{2}G_{F}\sum_{\alpha=1}^{2}
\sum_{I=1}^{6}x_{W\tilde{u}_{I}}\left (\Gamma^{d+}_{CL} \right )^{I}_{\alpha
j}\left [ \left (\Gamma^{d}_{CL} \right )^{\alpha 3}_{I}m_{b}
f_2(x_{\tilde{x}_{\alpha}^{-}\tilde{u}_{I}}) \right.
\nonumber \\
&  & \left. +\left (\Gamma^{d}_{CR}\right )^{\alpha3}_{I}
m_{\tilde{x}_{\alpha}^{-}}f_4(x_{\tilde{x}_{\alpha}^{-}\tilde{u}_{I}}) \right ],
\\
F^{\chi^{0}}_{1L}(0) &=&-\frac{G_F}{\sqrt{2}}
\sum_{\alpha=1}^{4}\sum_{I=1}^{6}x_{W\tilde{d}_{I}}
(\Gamma^{d+}_{NL})^{I}_{\alpha j} (\Gamma^{d}_{NL})^{\alpha 3}_{I}f_{7}
(x_{\tilde{x}_{\alpha}^{0}\tilde{d}_{I}}),
\\
F^{\chi^{0}}_{2R}(0) &=&-\sqrt{2}G_{F}\sum_{\alpha=1}^{4}\sum_{I=1}^{6}
x_{W\tilde{d}_{I}}(\Gamma^{d+}_{NL})^{I}_{\alpha
j}\left [ (\Gamma^{d}_{NL})^{\alpha
3}_{I}m_{b}f_2(x_{\tilde{x}_{\alpha}^{0}\tilde{d}_{I}}) \right.
\nonumber \\
&  & \left. +(\Gamma^{d}_{NR})^{\alpha
3}_{I}m_{\tilde{x}_{\alpha}^{0}}f_4(x_{\tilde{x}_{\alpha}^{0}\tilde{d}_{I}})
\right ], \label{eq:f2r}
\eeq
where $j=1$ for $b\to d$ and $j=2$ for $b\to s$
decay, respectively. $x_{ij}=m^{2}_{i}/m^{2}_{j}$ and $m_i$ is the
mass of the particle i. In our calculations, we have set $
m_{q}=0$ for $q=u,d,s$ since $m_{b}\gg m_u, m_d$ and $m_s$. The
one-loop integration functions $f_i(x)$ and the coupling constants
$\Gamma^{d}_{G(L,R)}, \Gamma^{d}_{C(L,R)}, \Gamma^{d}_{N(L,R)}$
which appear in $F_{1L}$ and $F_{2R}$ are listed in Appendix
\ref{app:fun}. Using the form factors in
Eqs.(\ref{eq:f1lh})-(\ref{eq:f2r}), we obtain the analytic
expressions for $C_k^{NP}(M_{W})$ and $C_{8g}^{NP}(M_{W})$. The
Wilson coefficient $C_{8g}^{NP}(M_{W})$ as given in
Eq.(\ref{eq:c8g}) is the same as that in Ref.\cite{tyy97} except
for some differences in expression. In Ref.\cite{tyy97}, the CKM
factor $-\lambda_t$ has not been extracted from Wilson
coefficients, and the CKM matrix elements have been absorbed into
the definition of the coupling constant $\Gamma^{U}_{L}$. See
Appendix \ref{app:fun} for more details.

For the effective vertex of the supersymmetric $b\to q \gamma$
penguin processes, we only consider it's contributions to
$C_{7\gamma}$. The explicit analytical expressions of the SUSY
contribution to $C_{7\gamma}$ induced by new particles
have been given in Ref.~\cite{tyy97}
\beq
\label{eq:c71}
C_{7\gamma}^{H^{-}}(m_W)&=&
 - \frac{1}{2}x_{th}\left \{ \cot ^2 \beta
 \left [ \frac{2}{3}f_1 (x_{th} ) + f_2 (x_{th} ) \right ]
   + \left [ \frac{2}{3}f_3 (x_{th} ) + f_4 (x_{th} ) \right ] \right \} \\
C_{7\gamma}^{\tilde{g}}(m_W)&=&
 - \frac{8}{9}\frac{{g_s^2 }}{{g^2 \lambda _t }}
 \sum\limits_{I = 1}^6 {x_{W\tilde d_I } } \left (\Gamma _{GL}^{d + } \right )_j^I
 \left [ \left (\Gamma _{GL}^d \right )_I^3 f_2 (x_{\tilde G\tilde d_I } )
 + \left (\Gamma _{GR}^d \right )_I^3
 \frac{{m_{\tilde G} }}{{m_b }}f_4 (x_{\tilde G\tilde d_I } ) \right ]
\\
C_{7\gamma}^{\tilde{\chi}^{-}}(m_W)&=&\frac{1}{\lambda_t}
\sum_{\alpha=1}^{2}\sum_{I=1}^{6}x_{W\tilde{u}_{I}}(\Gamma^{d+}_{CL})^{I}_{\alpha
j} \left \{ \left (\Gamma^{d}_{CL}\right )^{\alpha
3}_{I} \left [ f_1(x_{\tilde{x}_{\alpha}^{-}\tilde{u}_{I}})
+\frac{2}{3}f_2(x_{\tilde{x}_{\alpha}^{-}\tilde{u}_{I}}) \right ] \right.
\non
&\ \ & \left.
+\left (\Gamma^{d}_{CR}\right )^{\alpha3}_{I}\frac{m_{\tilde{x}_{\alpha}^{-}}}{m_b}
\left [ f_3(x_{\tilde{x}_{\alpha}^{-}\tilde{u}_{I}})
+\frac{2}{3}f_4(x_{\tilde{x}_{\alpha}^{-}\tilde{u}_{I}}) \right ] \right \}
\\
\label{eq:c72} C_{7\gamma}^{\tilde{\chi}^{0}}(m_W)&=&
-\frac{1}{3\lambda_t}\sum_{\alpha=1}^{4}\sum_{I=1}^{6}x_{W\tilde{d}_{I}}(\Gamma^{d+}_{NL})^{I}_{\alpha
j}\left [ \left (\Gamma^{d}_{NL}\right )^{\alpha
3}_{I}f_2(x_{\tilde{x}_{\alpha}^{0}\tilde{d}_{I}}) \right. \non
&  & \left.  + (\Gamma^{d}_{NR})^{\alpha
3}_{I}\frac{m_{\tilde{x}_{\alpha}^{0}}}{m_{b}}f_4(x_{\tilde{x}_{\alpha}^{0}\tilde{d}_{I}})
\right ].
\eeq

Now, we found all the supersymmetric contributions to the relevant Wilson
coefficients. We should remember that,
the only  source of  flavor violation in the mSUGRA model is the
usual CKM matrix in the SM. The flavor violation in the sfermion
sector at the electroweak scale is generated radiatively in the
mSUGRA model and consequently small. Therefore, If we
take the mixing matrices $\Gamma^U$ and $\Gamma^D$ as
given in the Appendix of Ref.\cite{ctfj}
\beq
\label{eq:ud}
\Gamma ^U  =
\left[ {\begin{array}{*{20}c}
   1 & 0 & 0 & 0 & 0 & 0  \\
   0 & 1 & 0 & 0 & 0 & 0  \\
   0 & 0 & {\cos \theta _{\tilde t} } & 0 & 0 & {\sin \theta _{\tilde t} }  \\
   0 & 0 & 0 & 1 & 0 & 0  \\
   0 & 0 & 0 & 0 & 1 & 0  \\
   0 & 0 & { - \sin \theta _{\tilde t} } & 0 & 0 & {\cos \theta _{\tilde t} }  \\
\end{array}} \right], \ \ \Gamma ^D  = \left[ {\begin{array}{*{20}c}
   1 & 0 & 0 & 0 & 0 & 0  \\
   0 & 1 & 0 & 0 & 0 & 0  \\
   0 & 0 & {\cos \theta _{\tilde b} } & 0 & 0 & {\sin \theta _{\tilde b} }  \\
   0 & 0 & 0 & 1 & 0 & 0  \\
   0 & 0 & 0 & 0 & 1 & 0  \\
   0 & 0 & { - \sin \theta _{\tilde b} } & 0 & 0 & {\cos \theta _{\tilde b} }  \\
\end{array}} \right]
\eeq
the gluino- and neutralino-mediated  diagrams
will not contribute to the decay processes considered here.
The new physics contributions will come from the charged-Higgs and
chargino diagrams only.

\subsection{$B\to PP$ decays in QCD factorization}

To calculate the decay amplitude  of the processes $B\to PP$, the
last but most important step is to calculate  hadronic  matrix
elements for the hadronization of the final-state quarks into
particular final states. At the present time, many approaches have
been put forward to settle the intractable problem. Such as the
native factorization \cite{bsw87}, the generalized factorization
\cite{ali9804,chen99}, the QCD FA \cite{bbns99,mgmc01,mm03} and the PQCD
approach \cite{pqcd}. In this paper, we employ the QCD FA
to calculate the branching ratios of $B \to PP$ decays.

In QCD FA, the contribution of the
non-perturbative sector is dominated in the form factors of $B\to
P$ transition and the nonfactorizable impact in  the hadronic
matrix elements is controlled by hard gluon exchange.  In the
heavy quark limit $m_b \gg \Lambda_{QCD}$ and to leading power in
$ \Lambda_{QCD}/m_{b}$, the hadronic matrix elements of the exclusive
 nonleptonic decays of the B meson into two light pseudoscalar
 mesons $P_1,P_2$ ($P_1$ absorbs the spectator quark coming from the B meson)
 can be written as \cite{bbns99}
\beq
\langle P_{1}P_{2}|O_i|B\rangle&=&\sum_{j}F_{j}^{B\rightarrow
P_{1}}\int_{0}^{1}dx
T_{ij}^{I}(x)\Phi_{P_2}(x)+(P_1\leftrightarrow P_2) \nonumber \\
&   &+\int_{0}^{1}d\xi\int_{0}^{1}dx\int_{0}^{1}dy
T_{i}^{II}(\xi,x,y)\Phi_{B}(\xi)\Phi_{P_1}(x)\Phi_{P_2}(y)
\label{eq:QCDF}
\eeq
where  $F_{j}^{B\rightarrow P_{1}}$ is the form factor describing
$B\to P_1$ decays. $T_{ij}^{I}$ and $T_{i}^{II}$ denote the
perturbative short-distance interactions and can be calculated by
the perturbation approach. $\Phi_{X}(x)(X=B,P_{1,2})$ are
the universal and nonperturbative
light-cone distribution amplitudes (LCDA) for B and $P_{1,2}$
meson respectively \cite{mgmc01}.
Weak annihilation effects are not included in Eq.(\ref{eq:QCDF}).

Consider the low energy effective Hamiltonian  Eq.(\ref{eq:eff})
and the unitary relation of the CKM matrix, the decay amplitude
can be written as
\beq \label{eq:aa}
{\cal A}(B\to P_{1}P_{2})=\frac{G_F}{\sqrt{2}}\sum_{p=u,c}
\sum_{i}V_{pb}V_{pq}^{*}C_i(\mu)\langle
P_{1}P_{2}|O_i|B\rangle
\eeq
the effective hadronic matrix
elements $\langle P_{1}P_{2}|O_i|B\rangle$ can be calculated by
employing the QCD factorization formula Eq.(\ref{eq:QCDF}).  When
considering order $\alpha_s$ corrections to the hard scattering
kernels $T_{ij}^{I}$ and $T_{i}^{II}$ from nonfactorizable single
gluon exchange vertex correction diagrams, penguin diagrams and
hard spectator scattering diagrams and the contributions from the
chirally enhanced power corrections \footnote{For more details of
various contributions and the corresponding Feynman loops,  see
for example Refs.~\cite{mgmc01,mm03} and references therein.},
Eq.(\ref{eq:aa}) can be rewritten  as\cite{du02}
\beq
\label{eq:af} {\cal A}^f(B\to P_{1}P_{2})=\frac{G_F}{\sqrt{2}}
\sum_{p=u,c}\sum_{i}V_{pb}V_{pq}^{*}a_i^p(\mu)\langle
P_{1}P_{2}|O_i|B\rangle _{F}
\eeq
Here $\langle P_{1}P_{2}|O_i|B\rangle _{F} $ is
the factorized matrix element
and can be factorizes into a form factor times a decay constant.
The explicit expressions for the decay amplitudes of $B\to P_1P_2$
can be found in Ref.\cite{ali9804}. For the processes involving
$\eta^{(')}$ in the final states, Ali {\it et al. } \cite{ali9804}
included the terms directly proportional to the so-called
charm decay constant  $f^c_{\eta^{(')}}$ of the $\eta^{(')}$ meson
in the decay amplitudes. We here ignored these terms because they
are very small in size. For the charmless B meson
decays considered here, the hadronic matrix elements $\langle
P_{1}P_{2}|O_{1,2}^{c}|B\rangle_{F}$ have no contributions.
Following Beneke {\sl et al.} \cite{mgmc01}, every coefficient
$a_i(P_1,P_2)$($i=1$ to $10$) can be split into two parts:
\beq
a_{i}(P_1,P_2)=a_{i,I}(P_1,P_2)+a_{i,II}(P_1,P_2)
\eeq
with
\begin{align}
\label{eq:ai}
a_{1,I}&=C_1+\frac{C_2}{N_c}[1+\frac{C_F\alpha_s}{4\pi}V_{P_2}],&
 a_{1,II}  &= \frac{{C_2 }}{{N_c }}\frac{{C_F \pi \alpha _s }}{{N_c }}H_{P_1 P_2 },
\nonumber\\
a_{2,I}&=C_2+\frac{C_1}{N_c}[1+\frac{C_F\alpha_s}{4\pi}V_{P_1}],&
a_{2,II}&  = \frac{{C_1 }}{{N_c }}\frac{{C_F \pi \alpha _s }}{{N_c }}H_{P_2 P_1 },
\nonumber\\
a_{3,I}&=C_3+\frac{C_4}{N_c}[1+\frac{C_F\alpha_s}{4\pi}V_{P_1}],&
a_{3,II}&  = \frac{{C_4 }}{{N_c }}\frac{{C_F \pi \alpha _s }}{{N_c }}H_{P_2 P_1 },
\nonumber\\
a_{4,I}^p & = C_4+\frac{{C_3 }}{{N_c }}[1+\frac{{C_F \alpha _s }}{{4\pi }}V_{p_2 } ]
+ \frac{{C_F \alpha _s }}{{4\pi }}\frac{{P_{p_2 ,2}^p }}{{N_c }},&
a_{4,II} & = \frac{{C_3 }}{{N_c }}\frac{{C_F \pi \alpha _s }}{{N_c }}H_{P_1 P_2 },
\nonumber\\
a_{5,I}&=C_5+\frac{C_6}{N_c}[1+\frac{C_F\alpha_s}{4\pi}(-V^{'}_{P_1})],&
a_{5,II}&  = \frac{{C_6 }}{{N_c }}\frac{{C_F \pi \alpha _s }}{{N_c }}(-H^{'}_{P_2 P_1 }),
\nonumber\\
a_{6,I}^p & = C_6+\frac{{C_5 }}{{N_c }}[1-6.\frac{{C_F \alpha _s }}{{4\pi }}]+\frac{{C_F
\alpha _s }}{{4\pi }}\frac{{P_{p_2 ,3}^p }}{{N_c }},&
a_{6,II} & = 0,
\nonumber\\
a_{7,I}&=C_7+\frac{C_8}{N_c}[1+\frac{C_F\alpha_s}{4\pi}(-V^{'}_{P_1})],&
a_{7,II} & = \frac{{C_8 }}{{N_c }}\frac{{C_F \pi \alpha _s }}{{N_c }}(-H^{'}_{P_2 P_1 }),
\nonumber\\
a_{8,I}^p  &=C_8+\frac{{C_7 }}{{N_c }}[1-6\frac{{C_F \alpha _s }}{{4\pi }}]+\frac{ \alpha
 _{em} }{{9\pi }}\frac{{P_{p_2 ,3}^{p,EW} }}{{N_c }},&
a_{8,II} &  = 0,
\nonumber\\
a_{9,I}&=C_9+\frac{C_{10}}{N_c}[1+\frac{C_F\alpha_s}{4\pi}V_{P_1}],&
a_{9,II} & = \frac{{C_{10} }}{{N_c }}\frac{{C_F \pi \alpha _s }}{{N_c }}H_{P_2 P_1 },
\nonumber\\
a_{10,I}^p & =C_{10}+\frac{{C_9 }}{{N_c }}[1+\frac{{C_F \alpha _s }}{{4\pi }}V_{P_2}]+
\frac{ \alpha _{em} }{{9\pi }}\frac{{P_{p_2 ,2}^{p,EW} }}{{N_c }},&
a_{10,II}&   = \frac{{C_{9} }}{{N_c }}\frac{{C_F \pi \alpha _s }}{{N_c }}H_{P_1 P_2 },
\end{align}
where $N_c =3$, $ C_F=4/3$, $a_{i,I}\equiv a_{i,I}(\mu)$ and $a_{i,II}\equiv
a_{i,II}(\mu_h)$ with $\mu\sim m_b$ and $\mu_h=\sqrt{\Lambda_h\mu}$ with
$\Lambda_h=0.5Gev$ as in Ref.\cite{mgmc01}. The terms $V_P^{(')}$
result from the vertex corrections, $H_{P_1,P_2}^{(')}$ describe the
hard-scattering spectator contributions,
$P_{P_2,2}^{p}$ and $ P_{P_2,3}^{p}$
 ($P_{P_2,2}^{p,EW}$  and $ P_{P_2,3}^{p,EW}$ ) arise from
the QCD (electroweak) penguin contributions and the contributions from
dipole operator $O_{8g}$ ( $O_{7\gamma}$ ). For the four penguin terms,
the subscript 2 or 3 indicates the twist of the
corresponding projection. The explicit expressions
of the functions $V_{P}^{(')},H_{P_1,P_2}^{(')}, P_{P_2,2}^{p}$,
$P_{P_2,3}^{p}$, $P_{P_2,2}^{p,EW}$ and $P_{P_2,3}^{p,EW}$
can be found in Ref.\cite{mgmc01}.

In QCD FA, the non-factorizable power-suppressed contributions are
neglected. However, the hard-scattering spectator interactions and
annihilation diagrams cannot be neglected because of the chiral
enhancement. Since they give rise to infrared endpoint
singularities when computed perturbatively, they can only be
estimated in a model-dependent way and with a large uncertainty.
In Refs.\cite{mgmc01,mm03} these contributions are parameterized
by two complex quantities, $X_H$ and $X_A$,
\beq X_{H,A} = \left (
1 + \rho_{H,A} e^{i\phi_{H,A}}\right ) \ln\frac{m_B}{\Lambda_h}
\eeq
where $\Lambda_h =0.5$ GeV, $\phi_{H,A}$ are free phases in
the range $[-180^\circ,180^\circ]$ and $\rho_{H,A}$ are real
parameters varying within $[0,1]$. In this paper, we use the
formulas as given in Ref.\cite{mgmc01} directly to estimate the
annihilation contributions to specific final state. Under the
convention of Ref.\cite{mgmc01}, the annihilation amplitude can be
written as
\beq
\label{eq:ann} {\cal A}^{ann}(B\to
P_1P_2)\propto\frac{G_F}{\sqrt{2}}\sum_{p=u,c}\sum_{i}V_{pb}V_{pq}^{*}
f_Bf_{P_1}f_{P_2}b_i(P_1,P_2)
\eeq
where $f_B$, $f_M$ are the
decay constants of B meson and final-state hadrons respectively.
The coefficients $b_i(P_1,P_2)$ describe the annihilation
contributions and generally depend on quantity $X_A$. For explicit
expressions of coefficients $b_i$ one can see Ref.\cite{mgmc01}.

Now the total decay amplitudes can be written as
\beq
{\cal A}(B\to P_1P_2)={\cal A}^{f}(B\to P_1P_2)+{\cal A}^{ann}(B\to P_1P_2)\, ,
\eeq
the corresponding branching ratio then takes the form
\begin{equation}
{\cal B}(B\to P_1P_2)=\tau_B\frac{|P_c|}{8\pi M_B^2}|{\cal A}(B\to P_1P_2)|^2\, ,
\end{equation}
where $\tau_B$ is the B meson lifetimes, and $ |P_c|$ is the absolute values of
two final-state hadrons' momentum in the B rest frame.
For the CP-conjugated decay modes, the branching ratios can be obtained by replacement
of $ \lambda_p \to \lambda_p^{*}$ in the expressions of decay amplitudes.

The new physics contributions to the branching ratios of $B \to
PP$ decays will be included by using the Wilson coefficients $C_i$
with the inclusion of the new physics parts as described in
Eqs.(\ref{eq:ck}-\ref{eq:c8g}) and (\ref{eq:c71}-\ref{eq:c72}).

\section{Numerical calculations} \label{sec:nc}

In this section, we first give the  input parameters needed in
numerical calculations, and then present the  numerical results and
make some theoretical analysis.

\subsection{input parameters}\label{sip}

\begin{itemize}
\item
The parameters $(A, \lambda, \bar{\rho},\bar{\eta})$ in Wolfenstein
parametrization of the CKM matrix. At present, the parameter $A$ and $\lambda$
have been well determined by experiments. In numerical calculation, we will
use $A=0.854$, $\lambda=0.2196$,
$\overline \rho=0.22\pm 0.10$, and $\overline \eta=0.35\pm 0.05$ as given in
Ref.\cite{pdg02}.

\item
Quark masses.
When calculating the decay amplitudes, the pole and current quark masses
will be used. For the former, we will use
$$m_u=4.2Mev,\ \ m_c=1.5Gev,\ \ m_t=175Gev,$$
$$m_d=7.6Mev,\ \ m_s=0.122Gev,\ \ m_b=4.62Gev .$$
The current quark mass depends on the renormalization scale.
In the $\overline{MS}$ scheme and at a scale of 2GeV, we fix
$$\overline {m }_u (2Gev)=2.4Mev, \ \ \overline {m}_d  (2Gev)=6Mev, $$
$$ \overline {m}_s  (2Gev) = 105Mev, \ \
\overline {m}_b  (\overline {m_b } ) = 4.26Gev,$$
as given in PDG 2002 \cite{pdg02}, and then employ the formulae
in Ref.\cite{gam96}
\beq
\overline m (\mu ) = \overline m (\mu _0 )\left [\frac{{\alpha _s (\mu )}}{{\alpha _s (\mu _0 )}}
\right]^{\frac{{\gamma _m^{(0)} }}{{2\beta _0 }}}
\left [1 + \left ( \frac{{\gamma _m^{(1)} }}{{2\beta _0 }}
- \frac{{\beta _1 \gamma _m^{(0)} }}{{2\beta _0^2 }} \right )
\frac{{\alpha _s (\mu ) - \alpha _s (\mu _0 )}}{{4\pi }} \right ]
\eeq
to obtain the current quark masses at any scale. The definitions of
$\alpha_s$, $\gamma _m^{(0)}$, $\gamma _m^{(1)}$, $\beta_0$, and $\beta_1$
can be found in Ref.\cite{gam96}.

\item
Form factors and decay constants. Following Ref.\cite{mm03}, we
also use
\beq F_0^{B\to \pi }(0)  = 0.28 \pm 0.05,
\quad  F_0^{B \to k }(0)  = 0.34 \pm0.05.
\eeq

The decay constants of $\pi$, $k$ and $B$ are \cite{mgmc01}
$$f_{\pi}=131Mev,\ \ f_k=160Mev,\ \ f_B=180Mev$$
For $\eta$ and $\eta^{'}$, mixing happens between them. The decay
constants of them can be parameterized by $f_q$, $f_s$ and the
mixing angle $\phi$ of $\eta-\eta^{'}$\cite{tpb98}
$$f_{\eta}^{u}=f_{\eta}^{d}=f_{\eta}^{q}cos\phi,\ \
f_{\eta}^{s}=-f_ssin\phi$$
$$f_{\eta^{'}}^{u}=f_{\eta^{'}}^{d}=f_{\eta}^{q}sin\phi,\ \
f_{\eta^{'}}^{s}=f_scos\phi$$
with
$$f_q=(1.07\pm0.02)f_{\pi}, \ \ f_s=(1.34\pm0.06)f_{\pi},\ \
\phi=(39.3\pm1.0)^{\circ}$$
Similarly, the form factors $F_0^{B \to \eta }(0) $ and $F_0^{B \to \eta' }(0) $
are parameterized as in Ref.\cite{tpb98}.

\item
For the parameters $\rho_{H,A}$ and $\phi_{H,A}$, we do not consider
the variation of these parameters but fix
$$
\rho_A=0.05, \ \ \phi_A=10^\circ, \ \ \rho_H=0, \ \ \phi_H=0^\circ
$$
in numerical calculation. For the parameter $\lambda_B$ appeared
in the B meson light-cone distribution amplitude, we also take
$\lambda_B=(350\pm 150)$ Mev as in Ref.\cite{mgmc01}.

\item
For the well-known $\pi, K, \eta^{(')}$ and $B$ meson masses, as well as
the B meson lifetimes, we use the values as given in Ref.\cite{pdg02}.

\item
The SUSY parameters at electroweak scale.
Within the parameter space still allowed by known constraints from the
data \cite{tyy97,hmdx98,hagiwara} ( such as the strong constraints from
the precise measurements of $Br(B \to X_s \gamma)$),
we choose two sets of SUSY parameters of the mSGURA model at the high
unification energy scale as listed in Table \ref{tab:mssm}.
The resulting masses of charged Higgs boson
and charginos obtained by using the program SUSPECT V 2.1 \cite{ajg02}
are also given in Table \ref{tab:mssm}.

\end{itemize}

In numerical calculations, we always use the cental values of
above input parameters unless explicitly stated otherwise.

\begin{table}[htb]
\doublerulesep 1.5pt
\caption{Two sets of SUSY parameters to be used in numerical calculation.
And the corresponding mass spectrum of charged Higgs boson and
the charginos. All masses are in unit of $GeV$.}
\label{tab:mssm}
\begin{center}
\begin{tabular} {c|c|c|c|c|c|c|c|c} \hline  \hline
Cases & $\tan\beta$ & $m_{\frac{1}{2}}$ & $m_0$ & $A_0$
&$sign(\mu)$ & $m_{H^\pm}$& $m_{\chi^\pm_1}$& $m_{\chi^\pm_2}$ \\
\hline
Case-A &$2$  & $300$ &$300$ &$0$    &$-$ & $782.3$ &247.0&595.9   \\ \hline
Case-B &$40$ &$150$ &$369$ &$-400$ &$+$ & $330.2$ &109.6&312.3  \\
\hline \hline
\end{tabular}
\end{center}
\end{table}

\subsection{Wilson coefficients: Case A and B }

From explicit calculations, we find that the SUSY corrections to
$B\to PP$ decays are mostly induced by the new physics parts of
the $C_{7\gamma}$ and $C_{8g}$, while the coefficients  $C_k^{NP}$
($k=3,4,5,6$) are indeed too small to modify their SM counterparts
effectively. The numerical results show that the
$C_{7\gamma}(m_b)$ and $C_{8g}(m_b)$ in mSUGRA model can be quite
different from that in the SM, and can even have the opposite sign
compared with their SM counterparts.

\subsubsection{Case A}

We firstly consider the Case A. For the SUSY part, since we take
the mixing matrix $\Gamma^{U}$ and $\Gamma^{D}$ as given in
Eq.(\ref{eq:ud}), the Feynman diagrams induced by the gluino and
neutralino exchanges do not contribute to the quark level decays
$b\to (s,d) \gamma$ and $b \to (s,d) g$. To a precision of
$\mathcal{O}(10^{-5})$, the SUSY contributions to $C_k(k=3,4,5,6)$
at the scale $m_W$ are the same for both $b\to s$ and $b\to d$
transitions. The contributions from the gauge boson $W^\pm$, the
charged Higgs and the charginos are
\beq \label{eq:ckc}
C_k^{SM}(m_W)&=&\{0.00155, -0.00197, 0.00066, -0.00197\}, \\
C_k^{H^\pm}(m_W)&=&\{-0.00001, 0.00004, -0.00001,0.00004\}, \\
C_k^{\tilde{\chi}^\pm}(m_w)&=&\{0, 0.00003,0, 0.00003\} .
\eeq
For $C_{7\gamma}(M_W)$ and $C_{8g}(M_W)$, the NLO level numerical results
are
\beq
C_{7\gamma } (m_W ) = \left\{ {\begin{array}{*{20}c}
   {\underbrace { - 0.2175}_{C_{7\gamma }^{SM} (m_W )}\underbrace
   { - 0.0422}_{C_{7\gamma }^{H^ \pm  } (m_W )}\underbrace
   { - 0.0007 - 0.0002I}_{C_{7\gamma }^{\tilde{\chi }^\pm } (m_W )} =-0.2604- 0.0002I,\ \ b \to d}  \\
   {\underbrace { - 0.2175}_{C_{7\gamma }^{SM} (m_W )}\underbrace
   { - 0.0422}_{C_{7\gamma }^{H^ \pm  } (m_W )}\underbrace
    {- 0.0009 - 0.0002I}_{C_{7\gamma }^{\tilde{\chi }^\pm   } (m_W )} =- 0.2606 - 0.0002I, \ \ b \to s}  \\
\end{array}} \right.
\eeq
\beq
\label{eq:c8gc}
C_{8g} (m_W ) = \left\{ {\begin{array}{*{20}c}
   {\underbrace { - 0.1178}_{C_{8g}^{SM} (m_W )}\underbrace
    { - 0.0473}_{C_{8g}^{H^ \pm  } (m_w )}\underbrace
    { - 0.0002 }_{C_{8g}^{\tilde{\chi }^\pm } (m_W )} = - 0.1653 , \ \ b \to d}  \\
   {\underbrace { - 0.1178}_{C_{8g}^{SM} (m_W )}\underbrace
    { - 0.0473}_{C_{8g}^{H^ \pm  } (m_w )}\underbrace
    { - 0.0002 - 0.0001I}_{C_{8g}^{\tilde{\chi }^\pm } (m_W )} =  - 0.1653 - 0.0001I, \ \ b \to s}  \\
\end{array}} \right.
\eeq
The new physics contributions to $C_k(M_W)$ ($k=3,4,5,6$) are clearly two orders
smaller than their SM counterparts and therefore can be neglected safely.
For $C_{7\gamma}(M_W)$ and $C_{8g}(M_W)$ the charged Higgs contribution
is dominant over the chargino contribution, but still much smaller than
their SM counterparts. Obviously the case A is not phenomenologically
interesting, since the SUSY effect is  too small to be separated from the SM
contribution though experimental measurements.

\subsubsection{Case B}

Now we turn to Case B. For this case, the SUSY contributions
to $C_k(k=3,4,5,6)$ are still negligibly small:
(a) the charged Higgs contributions are at the $\mathcal{O}$($10^{-7}$)
level; and (b) the chargino contributions are
at the $\mathcal{O}$($10^{-5}$) for both $b\to s$  and $b\to d$  transitions.

For $C_{7\gamma}(M_W)$ and $C_{8g}(M_W)$, however, the SUSY
contributions are significant:
\beq
\label{eq:c7gmwb}
C_{7\gamma } (m_W ) = \left\{ {\begin{array}{*{20}c}
   {\underbrace { - 0.2175}_{C_{7\gamma }^{SM} (m_W )}\underbrace
   { - 0.1128}_{C_{7\gamma }^{H^ \pm  } (m_W )}\underbrace
   { + 1.0111 + 0.0063I}_{C_{7\gamma }^{\tilde{\chi }^\pm } (m_W )} = 0.6808 + 0.0063I,\ \ b \to d}  \\
   {\underbrace { - 0.2175}_{C_{7\gamma }^{SM} (m_W )}\underbrace
   { - 0.1128}_{C_{7\gamma }^{H^ \pm  } (m_W )}\underbrace
    { + 1.0193 + 0.0091I}_{C_{7\gamma }^{\tilde{\chi }^\pm   } (m_W )} = 0.6890 + 0.0091I, \ \ b \to s}  \\
\end{array}} \right.
\eeq
\beq
\label{eq:c8gmwb}
C_{8g} (m_W ) = \left\{ {\begin{array}{*{20}c}
   {\underbrace { - 0.1178}_{C_{8g}^{SM} (m_W )}\underbrace
    { - 0.1103}_{C_{8g}^{H^ \pm  } (m_w )}\underbrace
    { + 0.4622 + 0.0007I}_{C_{8g}^{\tilde{\chi }^\pm } (m_W )} = 0.2341 + 0.0007I, \ \ b \to d}  \\
   {\underbrace { - 0.1178}_{C_{8g}^{SM} (m_W )}\underbrace
    { - 0.1103}_{C_{8g}^{H^ \pm  } (m_w )}\underbrace
    { + 0.4631 + 0.0010I}_{C_{8g}^{\tilde{\chi }^\pm } (m_W )} = 0.2350 + 0.0010I, \ \ b \to s}  \\
\end{array}} \right.
\eeq

At the lower scale $m_b$, they are
\begin{equation}
\label{eq:c7gmbb}
C_{7\gamma } (m_b ) = \left\{ {\begin{array}{*{20}c}
   {\underbrace { - 0.3067}_{C_{7\gamma }^{SM} (m_b )}\underbrace { + 0.5896 + 0.0039I}_{C_{7\gamma }^{H^ \pm  } (m_b )
   + C_{7\gamma }^{\tilde{\chi }^\pm} (m_b )} = 0.2829 + 0.0039I, \ \ b \to d}  \\
   {\underbrace { - 0.3067}_{C_{7\gamma }^{SM} (m_b )}\underbrace { + 0.5947 + 0.0058I}_{C_{7\gamma }^{H^ \pm  } (m_b )
   + C_{7\gamma }^{\tilde{\chi }^\pm} (m_b )} = 0.2880 + 0.0058I, \ \ b \to s}  \\
\end{array}} \right.
\end{equation}
\begin{equation}
\label{eq:c8gmbb}
C_{8g} (m_b ) = \left\{ {\begin{array}{*{20}c}
   {\underbrace { - 0.1500}_{C_{8g}^{SM} (m_b )}\underbrace { + 0.2449 + 0.0005I}_{C_{8g}^{H^ \pm  } (m_b )
   + C_{8g}^{\tilde{\chi }^\pm} (m_b )} = 0.0949 + 0.0005I, \ \ b \to d}  \\
   {\underbrace { - 0.1500}_{C_{8g}^{SM} (m_b )}\underbrace { + 0.2455 + 0.0007I}_{C_{8g}^{H^ \pm  } (m_b )
   + C_{8g}^{\tilde{\chi }^\pm} (m_b )} = 0.0955 + 0.0007I, \ \ b \to s}  \\
\end{array}} \right.
\end{equation}

From the numerical values in Eqs.(\ref{eq:c7gmwb}-\ref{eq:c8gmbb}),
one can see that
\begin{itemize}
\item
At the scale $M_W$, the charged Higgs contributions to both $C_{7\gamma}$
and $C_{8g}$ have the same sign with their SM counterparts, and are
comparable in size with them. The chargino contributions, however, have
an opposite sign with $C_{7\gamma}^{SM}$
and $C_{8g}^{SM}$, and  much larger in size than them.

\item
At both energy scales $m_W$ and $m_b$, the net new physics contributions
to $C_{7\gamma}$ and $C_{8g}$ are always positive and consequently
cancel their SM counterpart. The total value of these two
coefficients therefore become positive after the combination of the SM and
the new physics parts.

\item It is easy to understand why the new physics contributions
in case B are much larger than those in case A. In case A, the new
physics contributions from both charged Higgs boson and charginos
are negligibly small. In case B, however, we have much light
charged Higgs boson and charginos,  which leads to large SUSY
contributions. After the cancellation among the SM and SUSY contributions, the
net value of $C_{7\gamma}$ and $C_{8g}$ is positive.

\item
In case B, though $C_{7\gamma}(m_b)$ received a large
supersymmetric correction and has the opposite sign
with its SM counterpart, it's absolute value changes a little and
makes the theoretical prediction for the branching ratio of
$b\to s \gamma$ decay still consistent with the data.
The reason is rather simple:
the branching ratio  $Br(b\to s \gamma)$ is basically proportional
to $|C_{7\gamma}(m_b)|^2$.

\end{itemize}

\subsection{Branching ratios: data and theoretical prediction}

Using the decay amplitudes as given in Refs.\cite{ali98,du02} and the
coefficients $a_i$ in Eq.(\ref{eq:ai}), it is straightforward to calculate
the branching ratios of those twenty one $B \to PP$ decay modes
in the SM and mSUGRA model.

In order to show more details about the ways to include the SUSY contributions,
we present here, as an example,  the calculations
for the branching ratio $Br(B \to \pi^+ K^-)$.
The decay amplitudes of $B^0 \to \pi^+ K^-$ are
\beq
{ \cal A}^f (B^0 \to \pi^+ K^-)&=& -i\frac{G_F}{\sqrt{2}}
f_k F_0^{B\to\pi} \left ( m_B^2 -m_\pi^2\right ) \non
&& \times \left \{ V_{ub}V_{us}^* \left [ a_1 + a_4^u + a_{10}^u
+(a_6^u + a_8^u) R_4 \right ]
\right. \non
&& \left. + V_{cb}V_{cs}^*\left [ a_4^c + a_{10}^c +(a_6^c + a_8^c) R_4 \right ]
\right \}, \label{eq:afpik} \\
{ \cal A}^{ann}(B^0 \to \pi^+ K^-)&=& -i\frac{G_F}{\sqrt{2}}
f_B f_\pi f_k \left \{ -V_{tb}V_{ts}^* \left [ b_3(K\pi)
\frac{1}{2}b_3^{EW}(K,\pi)\right ] \right \}
\label{eq:annpik}
\eeq
with
\beq
R_4 = \frac{2m^2_K}{(m_b -m_u)(m_u+m_d)},
\eeq
where the coefficients $a_{i}^p$ have been given in Eq.(\ref{eq:ai}), the
coefficients $b_i(P_1,P_2)$ describe the annihilation contributions
\cite{mgmc01}.  Because of the strong Cabbibo suppression
($|V_{ub}V_{us}^*|^2 \propto \lambda^4$) on
the "tree" contribution ( the $a_1$ term ),
the four $B \to \pi K$ decays are QCD penguin dominant decay modes,
and strongly depend on ``large" coefficients $a_4^p$ and $a_6^p$.

We follow the same mechanism as described in Refs.\cite{ali9804,mgmc01}
to include the SUSY contributions to $B \to PP$ decays.

As mentioned previously, the SUSY contributions to the Wilson coefficients
of the 4-quark penguin operators are very small and have been neglected.
The large new magnetic penguin contributions in mSUGRA model
can manifest themselves as radiative corrections
to the Wilson coefficients $C_{4,6,8,10}$ (or equivalently to
$a_{j,I}^p$ with $j=4,6,8,10$ and $p=u,c$) and contained
in the quantities $P_{P_2,2}^{p}$, $P_{P_2,2}^{p,EW}$,
$ P_{P_2,3}^{p}$ and $ P_{P_2,3}^{p,EW}$.

For $B \to \pi^+ K^-$ decay, for example, the quantities $P_{K,2}^{p}$ and
$P_{K,2}^{p,EW}$ can be written as \cite{mgmc01}
\beq
P_{K,2}^p  &=& C_1 \left [\frac{4}{3}\ln \frac{{m_b }}{\mu }
+ \frac{2}{3} - G_k (s_p )\right ] + C_3
\left [\frac{8}{3}\ln \frac{{m_b }}{\mu } + \frac{4}{3} - G_K (0) - G_K (1)\right ]
\nonumber\\
 & & + \left (C_4  + C_6 \right )
 \left [\frac{{20}}{3}\ln \frac{{m_b }}{\mu } - 3G_K (0) - G_K (s_c ) - G_K (1)\right ]
\nonumber \\
&& - 6C_{8g}^{eff} \left ( 1+ \alpha_1^K + \alpha_2^K \right ),
\label{eq:pk2p}\\
P_{K,2}^{p,EW}&=&\left (C_1+N_cC_2\right )
\left [\frac{4}{3}\ln \frac{{m_b }}{\mu } + \frac{2}{3} - G_K (s_p )\right ]
- 9C_{7\gamma}^{eff} \left ( 1+ \alpha_1^K + \alpha_2^K \right ),
\label{eq:pk2pew}
\eeq
where $s_{u}=0$, $s_{c}=m_{c}^2/m_{b}^2$ are mass ratios involved in the
evaluation of penguin diagrams, $\alpha_1^K=0.3\pm 0.3$ and
$\alpha_2^K=0.1\pm 0.3$ are Gegenbauer moments for K meson \cite{mgmc01}.
$ C_{7\gamma }^{eff}  = C_{7\gamma }  - \frac{1}{3}C_5  - C_6$ and
$C_{8g}^{eff}  = C_{8g}  + C_5 $ are the so-called ``effective "
Wilson coefficients. The explicit expressions of
the functions $G_K(0)$, $G_K(1)$ and $G_K(s_p)$ can be found easily
in Ref.\cite{mgmc01}.
The twist-3 quantities $ P_{P_2,3}^{p}$ and $ P_{P_2,3}^{p,EW}$
receive the SUSY corrections in the same way as $ P_{P_2,2}^{p}$ and
$ P_{P_2,2}^{p,EW}$.

From Eqs.(\ref{eq:ai},\ref{eq:pk2p},\ref{eq:pk2pew}) and the numerical
results as listed in Table \ref{tab:a4}, one can see that
\begin{itemize}
\item
After the inclusion of SUSY contributions, the effective coefficients
$ C_{7\gamma }^{eff}$ and $C_{8g}^{eff}$ changed their sign from
negative to positive. The real parts of the coefficients
$a^p_{4,I}$ and $a^p_{6,I}$ are consequently changed by about
$60\%$ and $10\%$ respectively, but the imaginary parts of
$a^p_{j,I}$  remain unchanged.

\item
Since the magnitude of coefficients $a^p_{4,I}$ and $a^p_{6,I}$ is
larger than that of $a^p_{8,I}$ and $a^p_{10,I}$ by one or two orders,
the new physics contributions to $C_{8g}$ dominate the total
new physics corrections.

\item
Since only the coefficients $a^p_{j,I}$ for $j=4,6,8,10$ receive
the SUSY contributions, one naturally expect a moderate or large
new physics corrections to those penguin dominated B meson decays,
such as $B \to K\pi$ and $B \to K \eta'$ processes. The tree-dominated
decay modes, for example $B \to \pi\pi$ decays, remain basically
unaffected.

\end{itemize}

For the phenomenologically interesting $B \to K \eta'$ decays and other
penguin dominated decay modes studied here, the large
SUSY contributions will be included in the same way as for $B \to \pi^+ K^-$
decays.

\begin{table}[thb]
\caption{ The coefficients $C_{7\gamma}^{eff}(\mu)$,
$C_{8g}^{eff}(\mu)$ and $a^p_{j,I}$ ($j=4,6,8,10$ and $p=u,c$) for
$B \to \pi K$ decays in the SM and the case B of the mSUGRA model.} \label{tab:a4}
\begin{center}
\begin{tabular} {l|c|c|c|c|c|c } \hline  \hline
& \multicolumn{2}{|c|}{$\mu=m_b/2$} &
  \multicolumn{2}{|c|}{$\mu=m_b$} &
  \multicolumn{2}{|c }{$\mu= 2m_b$} \\ \cline{2-7}
 & SM & mSUGRA & SM & mSUGRA & SM & mSUGRA \\ \hline
$C_{7\gamma}^{eff}(\mu)$&$-0.276$ & $+0.221$
                         & $-0.270 $ & $+ 0.325$
                         & $-0.258$& $+0.422$ \\ \hline
$C_{8g}^{eff}(\mu)$& $-0.155$& $+0.058$
                         & $-0.142$ & $ +0.104 $
                         & $-0.130$ & $ +0.145$  \\ \hline
 $a^u_{4,I}\times 10^{3}$& $-24.3 -17.4i$ & $-41.7 -17.4 i$
                         & $-23.9 -14.4i$ & $-39.8 -14.4 i$
                         & $-22.6 -12.3i$ & $-37.4 -12.3 i$ \\ \hline
 $a^c_{4,I}\times 10^{3}$& $-31.4 -12.1i$ & $-48.8 -12.1 i$
                         & $-29.0 -10.4i$ & $-45.0 -10.4 i$
                         & $-26.9 - 9.0i$ & $-41.7 - 9.0 i$ \\ \hline
 $a^u_{6,I}\times 10^{3}$& $-54.0 -15.8i$ & $-58.3 -15.8 i$
                         & $-40.2 -13.6i$ & $-43.9 -13.6 i$
                         & $-32.3 -11.9i$ & $-35.7 -11.9 i$ \\ \hline
 $a^c_{6,I}\times 10^{3}$& $-60.4 - 2.9i$ & $-64.5 - 2.9 i$
                         & $-44.8 - 3.8i$ & $-49.0 - 3.8 i$
                         & $-36.0 - 4.0i$ & $-39.5 - 4.0 i$ \\ \hline \hline
 $a^u_{8,I}\times 10^{4}$& $ 4.7  - 0.6i$ & $ 3.3 -0.6 i$
                         & $ 3.0  - 1.1i$ & $ 1.4 -1.1 i$
                         & $ 1.8  - 1.4i$ & $-0.1 -1.4 i$ \\ \hline
 $a^c_{8,I}\times 10^{4}$& $ 4.6  - 0.3i$ & $ 3.2 -0.3 i$
                         & $ 2.7  - 0.5i$ & $ 1.1 -0.5 i$
                         & $ 1.4  - 0.6i$ & $-0.5 -0.6 i$ \\ \hline
$a^u_{10,I}\times 10^{4}$& $-12.0 +12.4i$ & $-18.0 +12.4 i$
                         & $-13.0 + 8.6i$ & $-19.9 +8.6 i$
                         & $-14.9 + 6.3i$ & $-23.0 +6.3 i$ \\ \hline
$a^c_{10,I}\times 10^{4}$& $-12.3 +12.5i$ & $-18.1 +12.5 i$
                         & $-13.4 + 8.9i$ & $-20.3 +8.9 i$
                         & $-15.4 + 6.6i$ & $-23.0 +6.6 i$ \\ \hline
\hline
\end{tabular}
\end{center}
\end{table}

Among twenty one $B \to PP$ decay modes considered here, twelve
of them have been measured so far. The individual measurements
and the world average for the branching ratios of these decays
\cite{hfag04} are shown in Table \ref{tab:exp}.

In Table \ref{tab:smmssm}, we show the theoretical predictions for
the CP-averaged branching ratios for $B\to PP$ decays in both SM
and the mSUGRA model (case B), assuming $\mu=m_b/2, m_b$ and $2
m_b$, respectively. And $Br^{f+a}$ and $Br^{f}$ denote the
branching ratios with or without the inclusion of annihilation
contributions, respectively. It is evident that some decay modes
have strong $\mu-$dependence, and the annihilation contributions
can also be significant for $B \to K \pi$ and $B \to K \eta'$
decays. In the following subsections, we present the numerical
results and show the dominant theoretical errors induced by the
uncertainties of input parameters, and focus on those well
measured decay channels.

\begin{table}[thb]
\caption{Experimental data of the branching
ratios for $B\to PP$ in unit of $10^{-6}$, taken from the HFAG
website \cite{hfag04}. For $\overline{B}^0 \to \overline{K}^0 \eta$
decay, the BaBar's result \cite{babar0311} will be used in our
analysis.} \label{tab:exp}
\begin{center}
\begin{tabular} {l|l|l|l|l} \hline  \hline
 Decay Modes & BaBar &  Belle  &   CLEO  &  Average  \\ \hline
$B^-\to\pi^-\overline{K}^0$ & $22.3\pm1.7\pm1.1$ & $ 22.0\pm1.9\pm1.1$
&$18.8_{ - 3.3 - 1.8}^{ + 3.7 + 2.1}$& $ 21.8\pm1.4 $
\\
$B^-\to\pi^0K^-$& $12.8_{-1.1}^{+1.2}\pm1.0$ & $ 12.0\pm1.3_{-0.9}^{+1.3}$
&$12.9_{ - 2.2-1.1}^{ + 2.4 + 1.2}$& $ 12.5_{-1.0}^{+1.1} $
\\
$\overline{B }^0\to\pi^+K^-$& $17.9\pm0.9\pm0.7$ & $ 18.5\pm1.0\pm0.7$
&$18.0_{ - 2.1-0.9}^{ + 2.3 + 1.2}$& $18.2\pm0.8 $
\\
$B^0\to\pi^0\overline{K}^0$& $11.4\pm1.7\pm0.8$ & $ 11.7\pm2.3_{-1.3}^{+1.2}$
&$12.8_{-3.3-1.4}^{ + 4.0 + 1.7}$& $11.7\pm1.4 $
\\ \hline
$\overline{B}^0\to\pi^{+}\pi^{-}$& $4.7\pm0.6\pm0.2$ & $ 4.4\pm0.6\pm0.3$
&$4.5_{ - 1.2 - 0.4}^{ + 1.4 + 0.5}$& $ 4.6\pm0.4 $
\\
$B^-\to\pi^-\pi^0$&$5.5_{ - 0.9}^{ + 1.0}\pm 0.6$& $5.0 \pm 1.2 \pm 0.5$
& $4.6_{ - 1.6 - 0.7}^{ + 1.8 + 0.6}$&$ 5.2 \pm 0.8$ \\
$\overline{B}^0\to\pi^0\pi^0$&  $2.1\pm0.6\pm0.3$ & $ 1.7\pm0.6\pm0.2$ & $<4.4$
&$1.9\pm0.5$
\\ \hline
$B^-\to K^-\eta$&  $3.4\pm0.8\pm0.2$ & $ 5.3_{-1.5}^{+1.8}\pm0.6$
& $2.2_{-2.2}^{+2.8}$ &$3.7\pm0.7$
\\
$B^-\to K^-\eta^{'}$ &  $76.9\pm3.5\pm4.4$ & $ 76\pm6\pm9$ & $80_{-9}^{+10}\pm7
$ & $77.6_{-4.5}^{+4.6}$
\\
$\overline{B}^0\to\overline{K}^0\eta$ &  $2.9\pm 1.0 \pm 0.2$ & $< 12 $& $<9.3$ &
$ 2.9\pm 1.0 \pm 0.2$ \\
$\overline{B }^0  \to \overline{K}^ 0 \eta^{'}$ &  $60.6\pm5.6\pm4.6$ & $68\pm10_{-8}^{+9}$
& $89\pm_{-16}^{+18}\pm9$ &$65.2_{-5.9}^{+6.0}$
\\ \hline
$B ^-\to\pi^-\eta$ &$5.3 \pm 1.0 \pm 0.3\ $& $5.4_{ - 1.7}^{ + 2.0}  \pm 0.6
$& $1.2_{ - 1.2}^{ + 2.8}$& $4.9_{ - 0.8}^{ + 0.9}$ \\
$B^-\to\pi^-\eta^{'}$&  $<4.5$ & $ <7$ & $<12$ &$<4.5$
\\
$\overline{B }^0  \to \pi ^ 0 \eta$ &  $<2.5$ & & $<2.9$ &$<2.5$
\\
$\overline{B}^0\to\pi^0\eta^{'}$  &  $<3.7$  && $<5.7$ &$<3.7$
\\  \hline
$\overline{B}^0\to\eta\eta$ &  $<2.8$ && $<18$ &$<2.8$
\\
$\overline{B}^0\to\eta\eta^{'}$ &  $<4.6$  &\ \ & $<27$ &$<4.6$
\\
$\overline{B}^0\to\eta^{'}\eta^{'}$ &  $<10$ &\ \ & $<47$ &$<10$
\\ \hline
$B^-\to K^-K^0$ &  $<2.5$ &  $<3.3$ &$<3.3$ &$<2.4$
\\
$\overline{B}^0\to\overline{K}^0K^{0}$ &  $<1.8$ & $<1.5 $& $<3.3$ &$<1.5$
\\
$\overline{B}^0\to K^+K^{-}$ &  $<0.6$ & $<0.7$& $<0.8$ &$<0.6$
\\ \hline\hline
\end{tabular}
\end{center}
\end{table}

\begin{table}[thb]
\doublerulesep 1.5pt
\caption{ The CP-averaged branching ratios of $B\to PP$ decays in the  SM
and minimal SUGRA model ( in unit of $10^{-6}$ ) by using  the
central values of input parameters. $Br^{f+a}$ and $Br^{f}$ denote
the results with or without the annihilation contributions. }
\label{tab:smmssm}
\begin{center}
\begin{tabular} {l|cc|cc|cc|cc|cc|cc} \hline  \hline
\multicolumn{1}{c|}{} & \multicolumn{4}{|c|}{$\mu=m_{b}/2$} &
\multicolumn{4}{|c|}{$\mu=m_{b}$} & \multicolumn{4}{|c}{$\mu=2m_{b}$}
 \\ \cline{2-13}
  Decays & \multicolumn{2}{|c|}{SM}& \multicolumn{2}{|c|}{mSUGRA} &
       \multicolumn{2}{|c|}{SM}& \multicolumn{2}{|c|}{mSUGRA} &
      \multicolumn{2}{|c|}{SM}& \multicolumn{2}{|c}{mSUGRA}
  \\ \cline{2-13}
\ \ &$Br^f$&$Br^{f+a}$ & $Br^f$ & $Br^{f+a}$ & $Br^f$&$Br^{f+a}$
    &$Br^f$&$Br^{f+a}$ & $Br^f$ & $Br^{f+a}$ & $Br^f$&$Br^{f+a}$
\\ \hline \hline
$B^-\to\pi^-\overline{K}^0$ &$13.6$&$16.2$&$20.5$&$23.6$&$12.7$&$14.7$&$19.1$&$21.6$&
12.0&13.6&18.1&20.1 \\
$B^-\to\pi^0K^-$&8.1&9.3&11.8&13.4&7.6&8.6&11.2&12.4&7.3&8.1&10.7&11.7
\\
$\overline{B }^0\to\pi^+K^-$&10.5&12.4&16.3&18.8&10.0&11.6&15.6&17.5&9.7&10.9&15.1&16.7
\\
$
B^0\to\pi^0\overline{K}^0$&4.4&5.3&7.1&8.3&4.1&4.8&6.6&7.6&3.9&4.5&6.3&7.0\\
\hline
$\overline{B}^0\to\pi^+\pi^-$&9.0&9.6&9.2&9.9&8.8&9.3&9.0&9.5&8.6&9.0&8.8&9.3
\\
$ B^-\to\pi^-\pi^0$&6.1&-&6.1&-&6.3&-&6.3&-&6.4&-&6.4&- \\
$ \overline{B}^0\to\pi^0\pi^0$&0.35&0.39&0.43&0.47&0.31&0.32&0.37&0.39&0.31&0.31&0.36&0.37
\\ \hline
$B^-\to K^-\eta$&2.7&2.7&3.6&3.6&2.7&2.7&3.6&3.6&2.5&2.6&3.4&3.5
\\
$B^-\to K^-\eta^{'}$&36.6&47.6&48.1&60.6&30.0&38.1&40.2&49.4&26.7&32.8&36.2&43.3
\\
$\overline{B}^0\to\overline{K}^0\eta$
&2.1&2.1&2.9&2.9&2.0&2.0&2.8&2.8&1.9&1.9&2.7&2.7\\
$\overline{B }^0  \to \overline{K}^ 0 \eta^{'}$
&34.6&44.5&45.1&56.3&28.3&35.4&37.5&45.7&24.9&30.3&33.8&39.8
\\ \hline
$B^-\to\pi^-\eta$&4.53&4.46&4.82&4.76&4.42&4.38&4.70&4.68&4.44&4.42&4.73&4.72
\\
$B^-\to\pi^-\eta^{'}$&3.98&3.97&4.17&4.17&3.73&3.73&3.91&3.92&3.69&3.69&3.86&3.86
\\
$\overline{B}^0\to\pi^0\eta$&0.33&0.32&0.43&0.42&0.29&0.29&0.39&0.39&0.28&0.28&0.37&0.37
\\
$\overline{B}^0\to\pi^0\eta^{'}$&0.32&0.32&0.38&0.39&0.25&0.25&0.31&0.31&0.22&0.22&0.27&0.27
\\ \hline
$$$$$$$$$$$$$$$$$$
$\overline{B}^0\to\eta\eta$&0.20&0.29&0.23&0.33&0.19&0.27&0.23&0.31&0.21&0.27&0.24&0.31
\\
$\overline{B}^0\to\eta\eta^{'}$&0.35&0.42&0.39&0.46&0.32&0.37&0.36&0.41&0.33&0.37&0.37&0.41
\\
$\overline{B}^0\to\eta^{'}\eta^{'}$ &0.18&0.30&0.20&0.32&0.16&0.25&0.17&0.27&0.16&0.24&0.17&0.26
\\ \hline
$B^-\to
K^-K^0$&0.69&0.82&0.96&1.11&0.63&0.73&0.88&1.00&0.58&0.66&0.83&0.92
\\
$\overline{B}^0\to\overline{K}^0K^{0}$&0.63&0.83&0.88&1.11&0.57&0.72&0.80&0.98&0.53&0.64&0.75&0.89
\\
$\overline{B}^0\to K^+K^{-}$& - &0.06& - &0.06& - &0.03& - &0.03&-&0.02&-&0.02
\\         \hline\hline
\end{tabular}
\end{center}
\end{table}

\subsection{$B \to \pi \pi$ and $K\pi$ decays}

The three $B\to\pi\pi$ decays are tree-dominated decay modes. The
central values and the major errors of the branching ratios (in
units of $10^{-6}$) in the SM and mSUGRA model are \beq
Br(\overline{B}^0 \to \pi^+ \pi^-)&=& \left \{\begin{array}{ll}
 9.3 \pm 0.3 (\mu) \; ^{+3.7}_{-3.1} (F_0) \; ^{+0.9}_{-1.6}(\gamma)
 & {\rm in \ \ SM}, \\
 9.5 \pm 0.3 (\mu) \; ^{+3.8}_{-3.2} (F_0)\;  ^{+1.0}_{-1.8}(\gamma)
 & {\rm in \ \ mSUGRA}, \\
\end{array} \right. \label{eq:pipi1} \\
Br(B^- \to \pi^- \pi^0)&=& \left \{\begin{array}{ll}
 6.3\; ^{+0.2}_{-0.1}(\mu)\; ^{+2.2}_{-1.9} (F_0)
\; ^{+0.0}_{-0.2}(\gamma)
 & {\rm in \ \ SM}, \\
 6.3\; ^{+0.2}_{-0.1} (\mu)\; ^{+2.2}_{-1.9} (F_0)
\; ^{+0.0}_{-0.2}(\gamma)
& {\rm in \ \ mSUGRA}, \\
\end{array} \right. \label{eq:pipi2} \\
Br(\overline{B}^0 \to \pi^0 \pi^0)&=& \left \{\begin{array}{ll}
 0.32\; ^{+0.07}_{-0.01}(\mu)\; ^{+0.09}_{-0.08} (F_0)
\; ^{+0.14}_{-0.11}(\gamma)
 & {\rm in \ \ SM}, \\
 0.39\; ^{+0.08}_{-0.02} (\mu)\; ^{+0.12}_{-0.10} (F_0)
\; ^{+0.18}_{-0.14}(\gamma)
 & {\rm in \ \ mSUGRA}, \\
\end{array} \right. \label{eq:pipi3}
\eeq
where the three major errors are induced by the uncertainties
$m_b/2 \leq \mu \leq 2 m_b$, $F_0^{B\to \pi} = 0.28 \pm 0.05$ and $
\gamma= 60^\circ \pm 20^\circ$.

\begin{figure}[thb]
\centerline{\mbox{\epsfxsize=7cm\epsffile{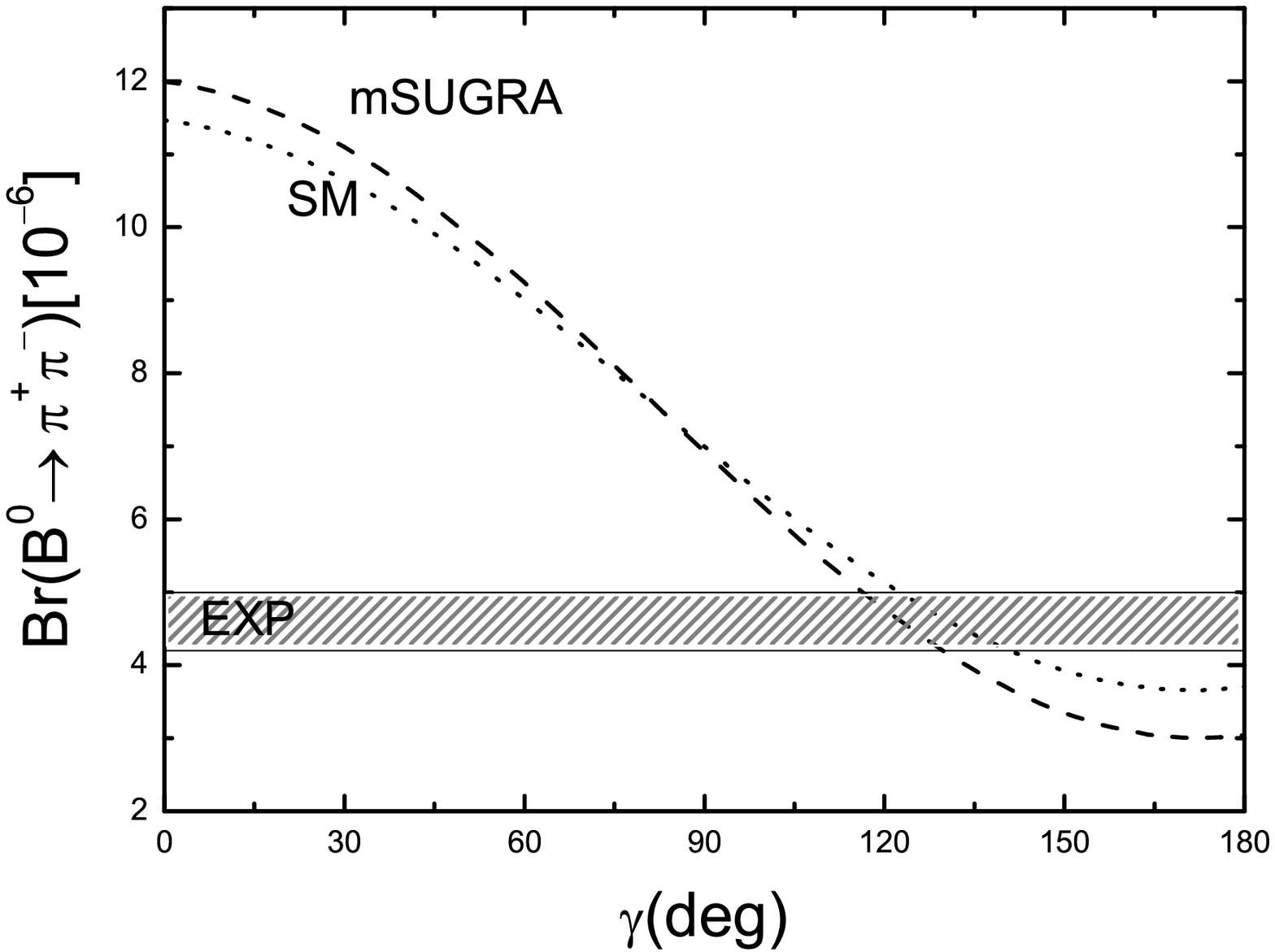}
\epsfxsize=7cm\epsffile{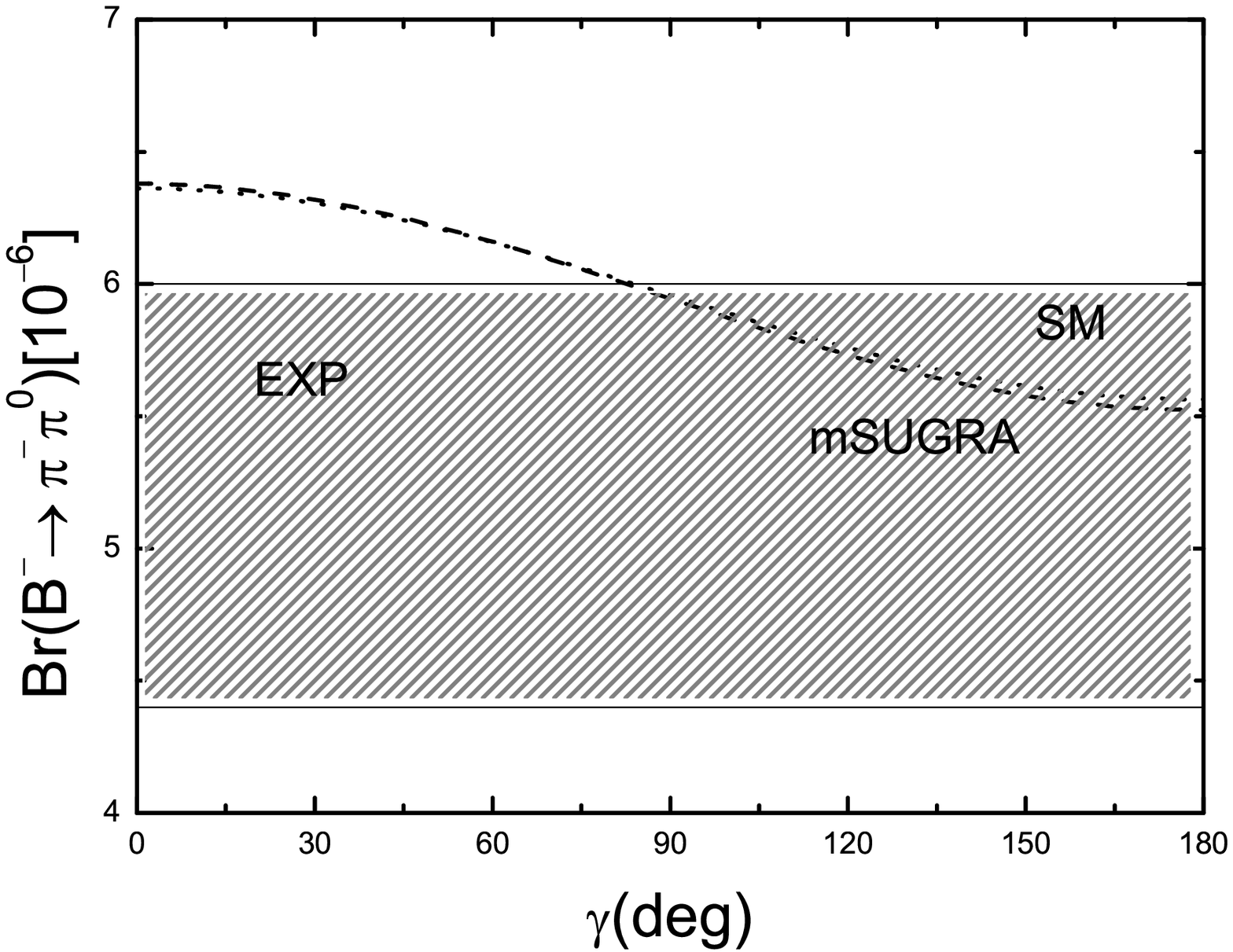}}}
\vspace{0.1cm}
\epsfxsize=7cm\epsffile{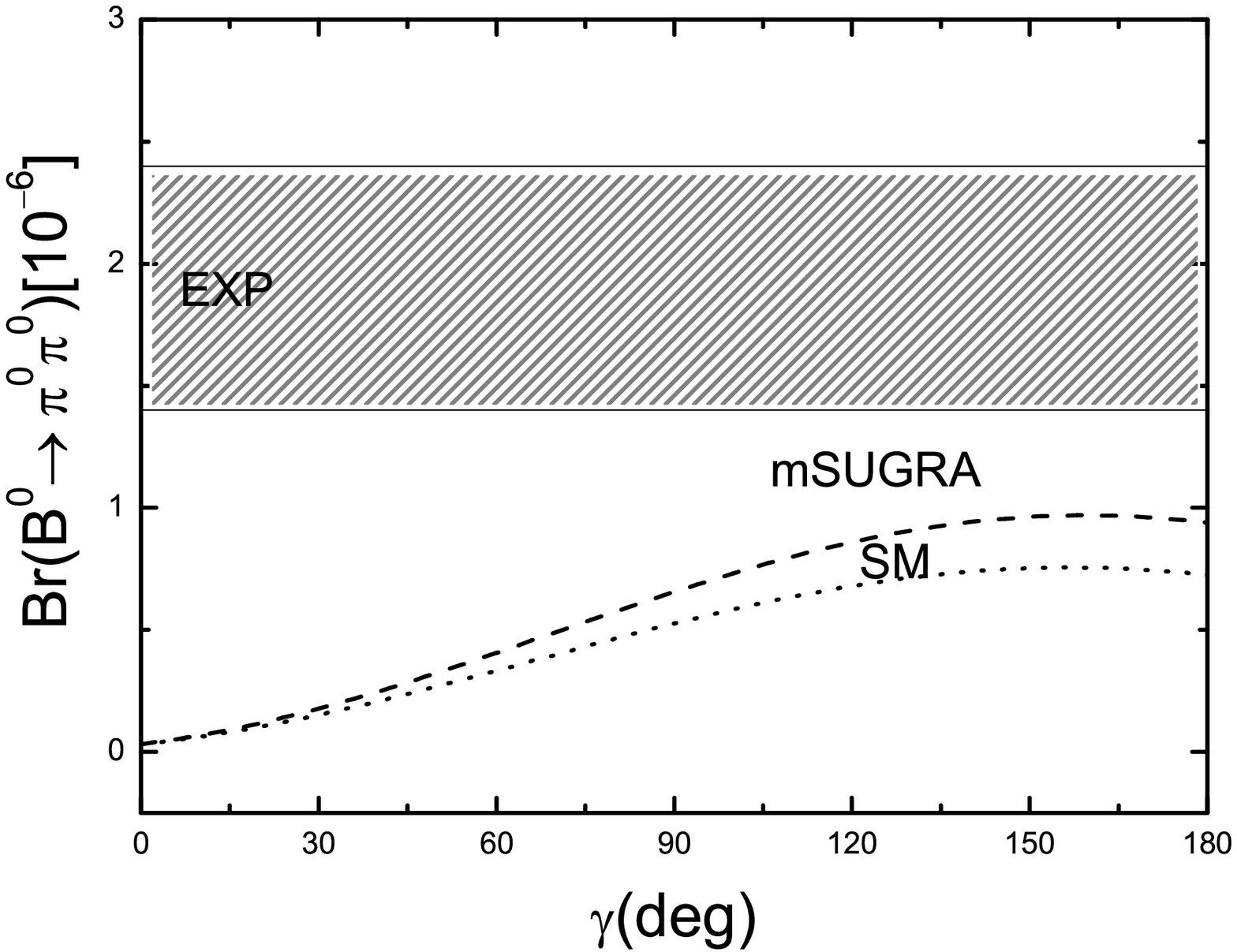}
\vspace{0.1cm}
\caption{The $\gamma$ dependence of the branching ratios of $B \to \pi\pi$
decays in the SM and  mSUGRA model.
The dots and solid curves show the central values of the SM and mSUGRA
predictions. The horizontal gray bands show the corresponding experimental
measurements as given in Table III.}
\label{fig:pipi}
\end{figure}

Fig.~\ref{fig:pipi} shows the $\gamma$ dependence of the
branching ratios for three $B \to \pi\pi$ decays. The dots and
dashed curves correspond to the central values of the theoretical
prediction in the SM and mSUGRA model \footnote{The central values
of all input parameters except for the CKM angle $\gamma$ are used
in this and other similar figures. The theoretical uncertainties
are not shown in all such kinds of figures.}, respectively. The
horizontal slashed bands show the data as given in Table
\ref{tab:exp}.

\begin{figure}[thb]
\centerline{\mbox{
\epsfxsize=8cm\epsffile{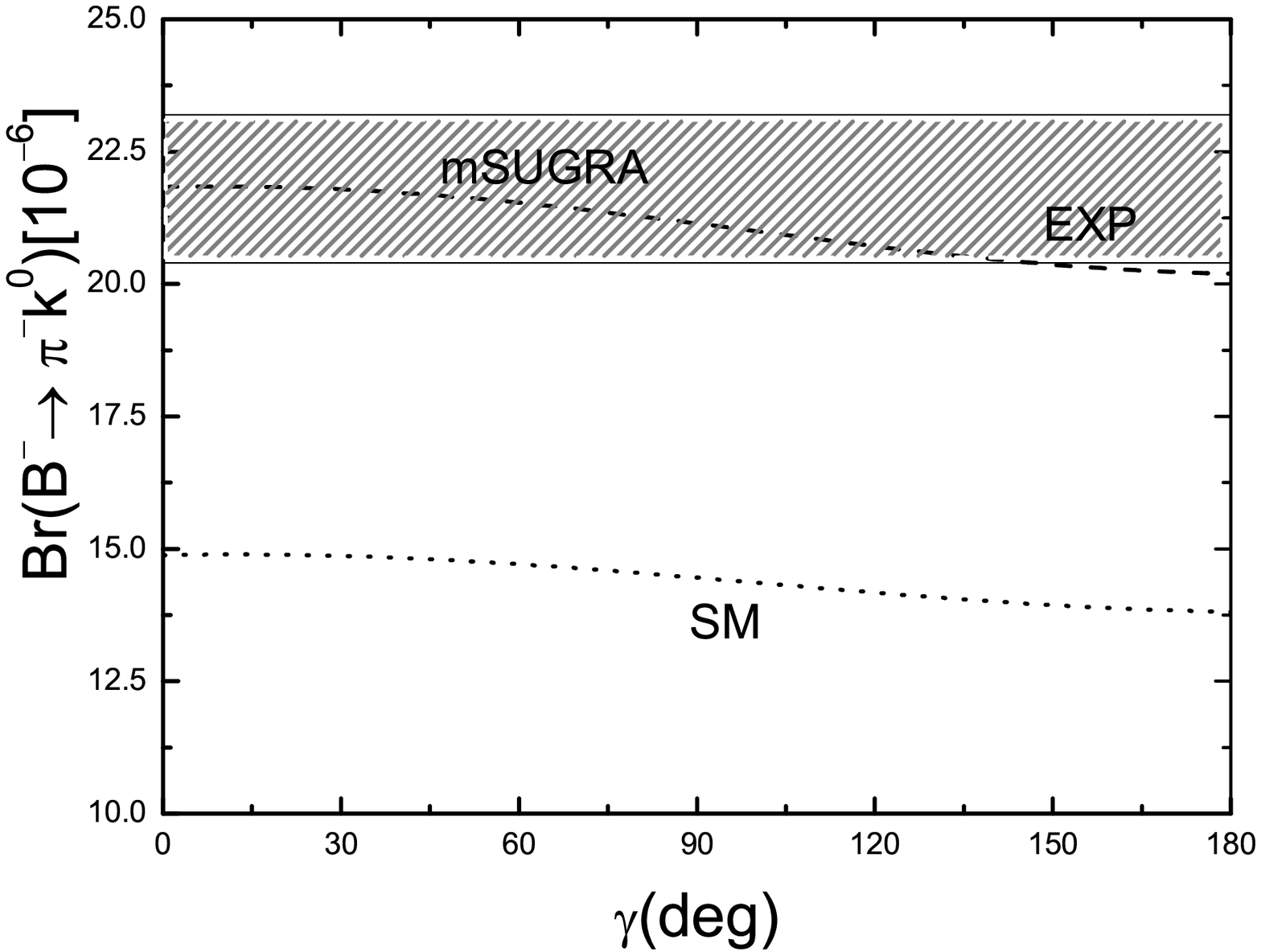}
\epsfxsize=8cm\epsffile{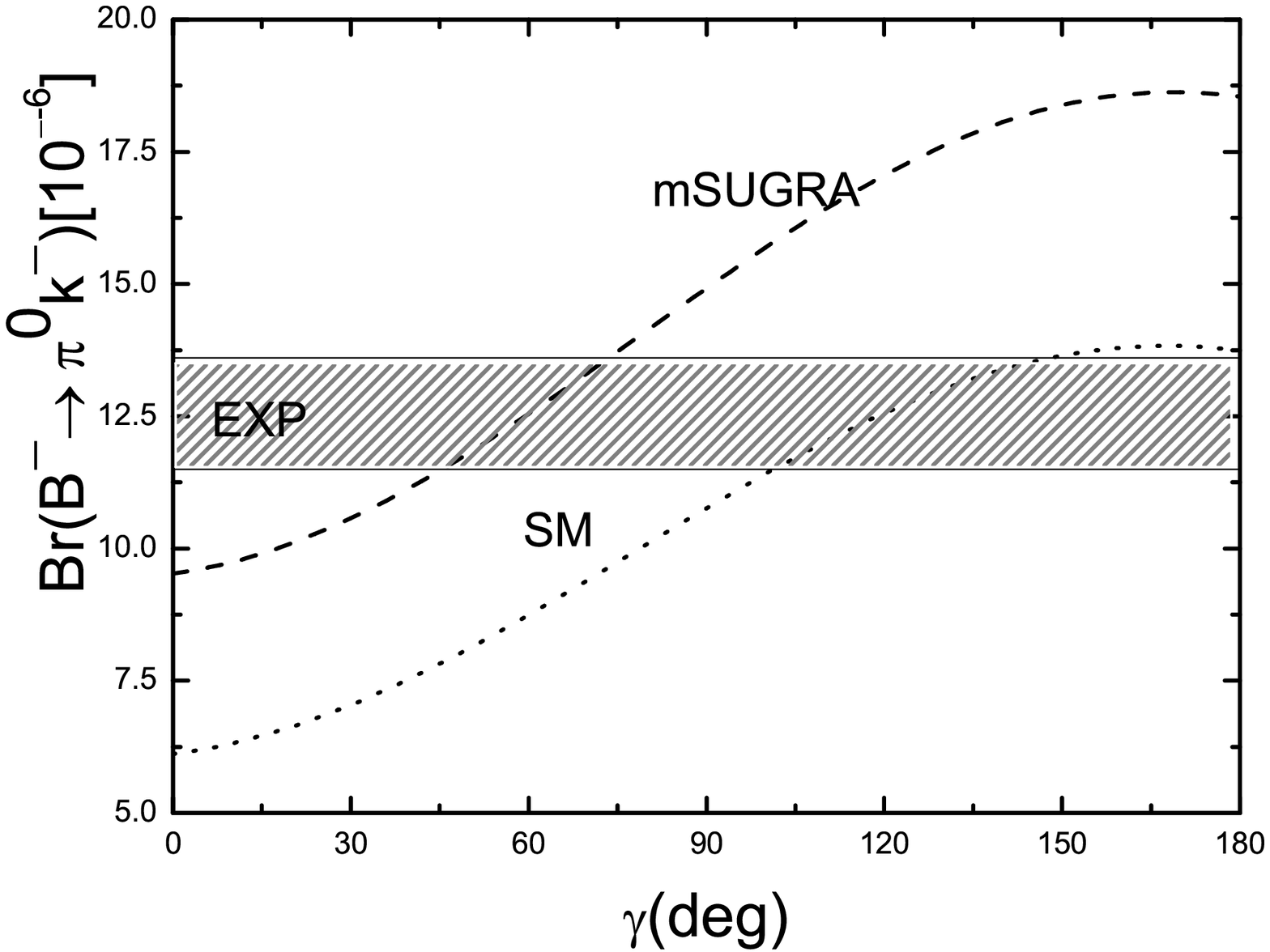}}}
\vspace{0.1cm}
\centerline{\mbox{
\epsfxsize=8cm\epsffile{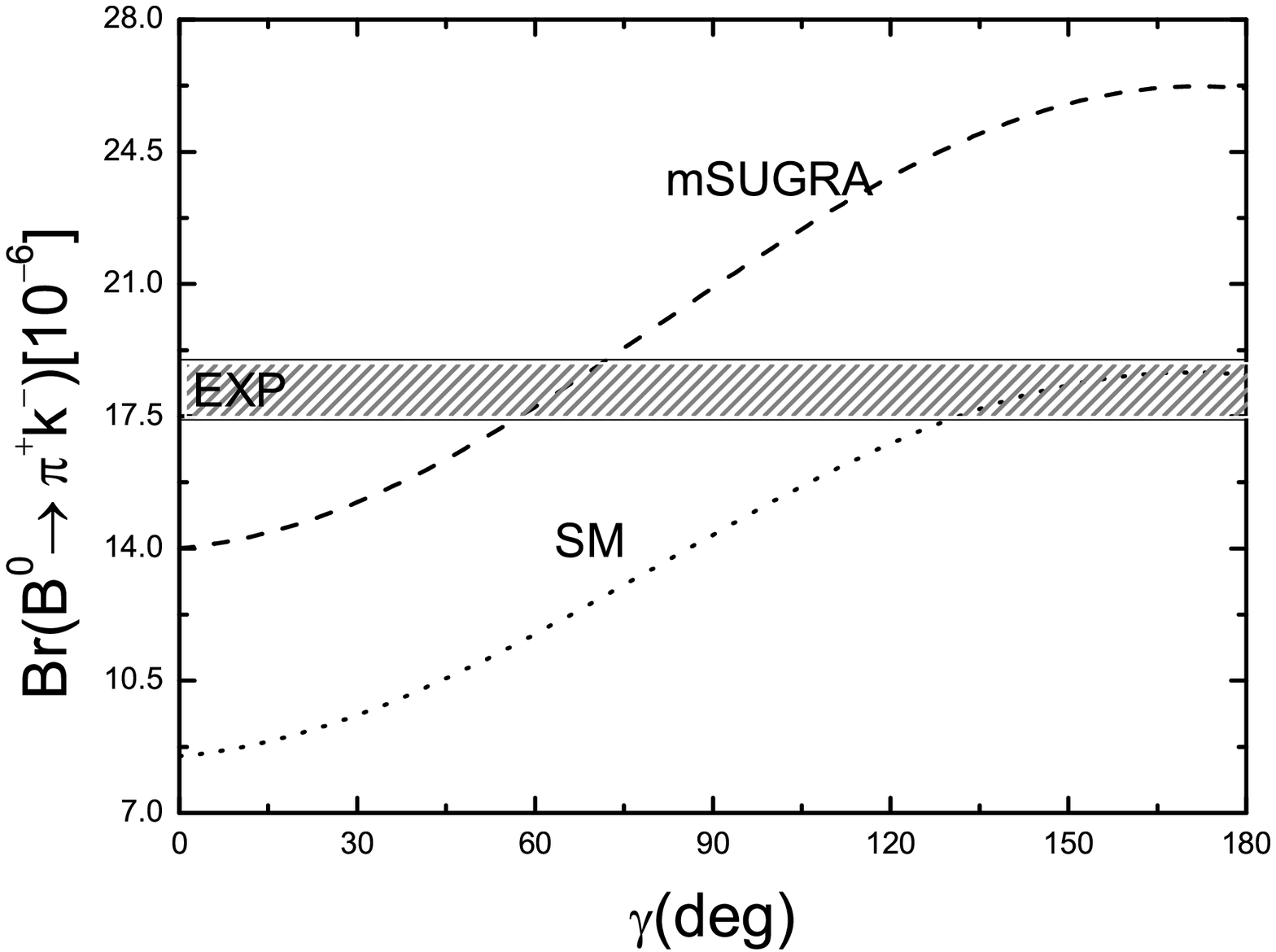}
\epsfxsize=8cm\epsffile{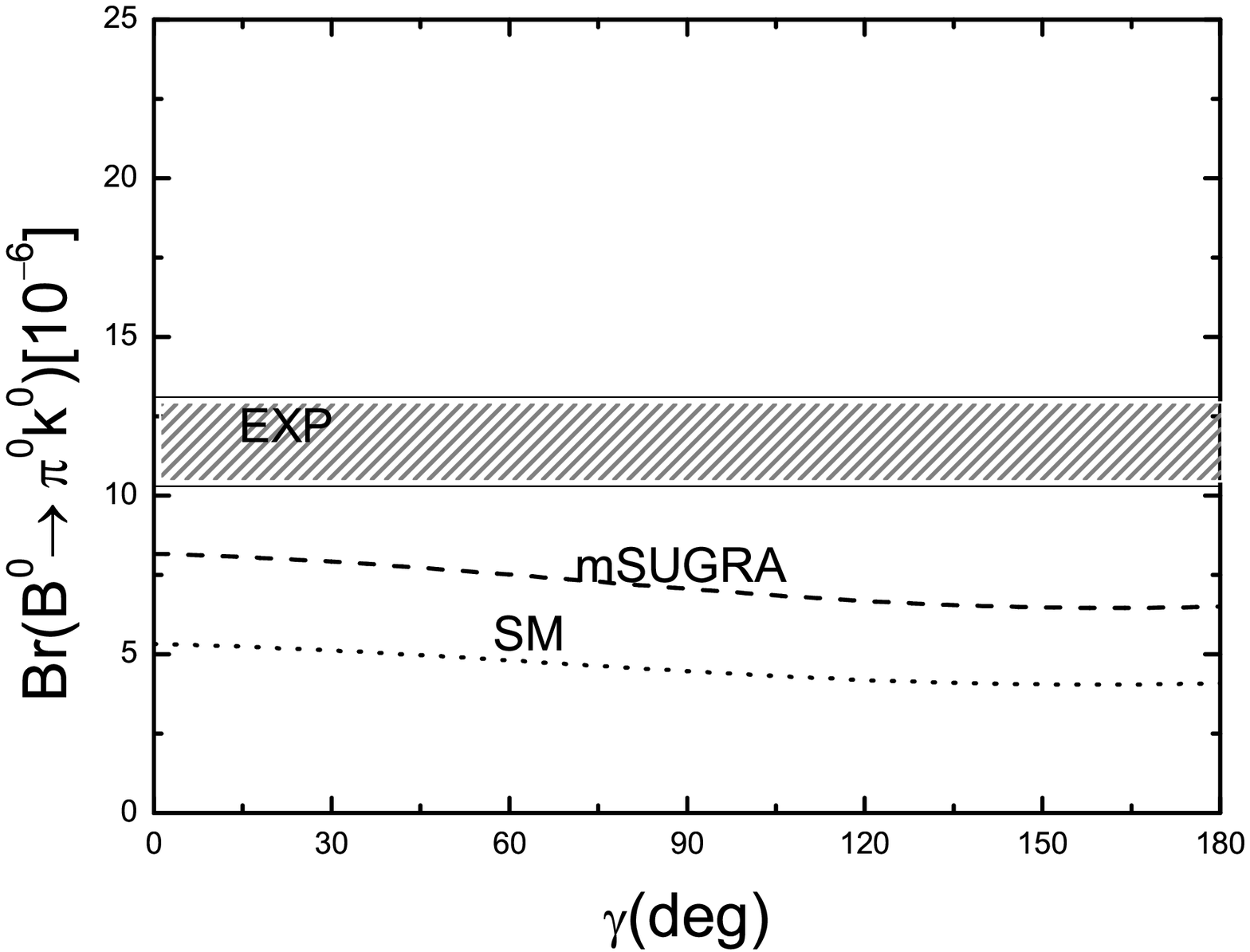}}}
\vspace{0.1cm}
\caption{The $\gamma$ dependence of the branching ratios of the four
$B \to K \pi$ decays in the SM and  minimal SUGRA model.
The dots and solid curves show the central values of the SM and mSUGRA
predictions. The horizontal gray bands show the corresponding experimental
measurements as given in Table III.}.
\label{fig:kpi}
\end{figure}

From Fig.~\ref{fig:pipi} and the numerical results as given in
Eqs.(\ref{eq:pipi1}-\ref{eq:pipi3}), one can see that
\begin{itemize}

\item
For these tree-dominated decays, the SUSY corrections considered here
are very small.

\item The theoretical predictions  strongly depend on the value of
the form factor $F_0^{B\to \pi}$.

\item For $B^0 \to \pi^0 \pi^0$ decay, the theoretical
prediction in QCD factorization is about five times smaller than
the measured value and cannot become consistent with the data
within the whole parameter space.

\item The central value of $Br(B^0 \to \pi^+ \pi^-)$ is much
larger than the data, but can become consistent with the data if
one uses a smaller form factor $F_0^{B\to \pi}$ or a large angle
$\gamma \sim 120^\circ$. But a small $F_0^{B \to \pi}$ is
disfavored by the large measured decay rates for $B \to \pi^0
\pi^0$ and $K \pi$ decay modes, while a large $\gamma$ around
$120^\circ$ is also in conflict with the global fit result $40^0 <
\gamma < 78^\circ$ at $95\%$C.L.\cite{ckmfit} and the latest
direct experimental measurement $\gamma = 81^\circ \pm
19^\circ(stat.) \pm 13^\circ (sys.) \pm 11^\circ (model)$
\cite{abe04}.

\end{itemize}

In the SM, the four $B \to K \pi$ decays are dominated by the
$b\to s g$ gluonic penguin diagrams, with additional contributions
from $b \to u$ tree and electroweak penguin diagrams. For these
decay modes, although the SM predictions can become consistent
with the measured values after considering the still large
theoretical uncertainties, but the central values of the SM
prediction are  indeed much smaller than the measured values even
after the inclusion of annihilation contributions. In the mSUGRA
model, the new penguin diagrams induced by new particles can
contribute effectively to $B \to K \pi$ decays. The numerical
results (in unit of $10^{-6}$)  are
\beq Br(B^- \to \pi^-
\overline{K}^0)&=& \left \{\begin{array}{ll}
 14.7 \; ^{+1.4}_{-1.1}(\mu) \; ^{+5.3}_{-4.8} (F_0)
\; ^{+0.1}_{-0.2} (\gamma) \; ^{+4.5}_{-2.8} (\bar{m}_s)
 & {\rm in \ \ SM}, \\
 21.6 \; ^{+2.0}_{-1.5} (\mu) \; ^{+7.9}_{-6.6} (F_0)
\; ^{+0.2}_{-0.3} (\gamma)\; ^{+5.8}_{-3.6} (\bar{m}_s)
 & {\rm in \ \ mSUGRA}, \\
\end{array} \right. \label{eq:pik1} \\
Br(B^- \to \pi^0 K^-)&=& \left \{\begin{array}{ll} 8.6 \;
^{+0.7}_{-0.5}(\mu) \; ^{+3.4}_{-2.6} (F_0) \; ^{+1.5}_{-1.1}
(\gamma) \; ^{+2.3}_{-1.4} (\bar{m}_s)
 & {\rm in \ \ SM}, \\
 12.4 \; ^{+1.0}_{-0.7} (\mu) \; ^{+4.4}_{-3.8} (F_0)
\; ^{+1.7}_{-1.2} (\gamma) \; ^{+3.0}_{-1.9} (\bar{m}_s)
& {\rm in \ \ mSUGRA}, \\
\end{array} \right. \label{eq:pik2}\\
Br(\overline{B}^0 \to \pi^+ K^-)&=& \left \{\begin{array}{ll}
 11.6 \; ^{+0.9}_{-0.7}(\mu) \; ^{+4.2}_{-3.5} (F_0)
\; ^{+1.9}_{-1.3} (\gamma) \; ^{+3.0}_{-1.9} (\bar{m}_s)
 & {\rm in \ \ SM}, \\
 17.5 \; ^{+1.3}_{-0.9} (\mu) \; ^{+6.4}_{-5.4} (F_0)
\; ^{+2.3}_{-1.6} (\gamma) \; ^{+4.8}_{-3.0} (\bar{m}_s)
 & {\rm in \ \ mSUGRA}, \\
\end{array} \right. \label{eq:pik3} \\
Br(B^0 \to \pi^0 \overline{K}^0)&=& \left \{\begin{array}{ll}
 4.8 \; ^{+0.5}_{-0.4}(\mu) \; ^{+1.8}_{-1.5} (F_0)
\; ^{+0.2}_{-0.3} (\gamma) \; ^{+1.8}_{-1.1} (\bar{m}_s)
 & {\rm in \ \ SM}, \\
 7.6 \; ^{+0.7}_{-0.5} (\mu) \; ^{+2.8}_{-2.6} (F_0)
 \; \pm 0.3  (\gamma) \; ^{+2.3}_{-1.4} (\bar{m}_s)
 & {\rm in \ \ mSUGRA}, \\
\end{array} \right. \label{eq:pik4}
\eeq where the second and fourth error are induced by the
uncertainties  $F_0^{B\to \pi} = 0.28 \pm 0.05$, $F_0^{B\to K} =
0.34 \pm 0.05$,   and $\overline{m}_s(2 Gev)=(105\pm20) Mev$.

Fig.~\ref{fig:kpi} shows the $\gamma$ dependence of the
branching ratios for four  $B \to K\pi$ decays. The dots and
dashed curves correspond to the central values of the theoretical
prediction in the SM and mSUGRA model, respectively. The
horizontal slashed bands show the data as given in Table
\ref{tab:exp}.

From Fig.~\ref{fig:kpi} and the numerical results as given in
Eqs.(\ref{eq:pik1}-\ref{eq:pik4}),
one can see that the SUSY contributions can provide  $\sim 50\%$
enhancement to the corresponding branching ratios, and such
enhancements can improve the consistency between the
theoretical predictions and the data effectively. The central values of
the theoretical predictions for $Br(B \to K \pi)$ in the mSUGRA model
become well consistent with the experimental measurements.

As for the ratio $R_n$ and $R_c$ as defined in section I,
the SM relation $R_c\approx R_n$ remain unchanged in the mSUGRA model.
The central values of these two ratios are:
\beq
R_c^{SM} &=& 1.17, \quad  R_c^{mSUGRA} = 1.15, \\
R_n^{SM} &=& 1.20, \quad  R_n^{mSUGRA} = 1.16.
\eeq
The reason is that the SUSY contributions to the four
$B \to K \pi$ decays are similar in nature, and thus cancelled in the
ratio of the corresponding branching ratios.

\subsection{$B \to K \eta^{(')}$ decays}

The unexpectedly large branching ratios of $Br( B^\pm \to K^\pm
\eta')$ and $Br(B^0 \to K^0 \eta')$ were reported by
CLEO, BaBar and Belle
Collaborations\cite{cleo03,babar03,belle03,hfag04}, and have been
studied in the SM \cite{as97} and new physics models by many
authors\cite{etapnp,etapm3}.

For the branching ratios of $B \to K^\pm \eta'$ and $K^0
\eta'$ decays, as can be seen from table \ref{tab:exp} and
\ref{tab:smmssm}, the experimental measurements are about twice
that of the the central values of the SM predictions in QCD
factorization. In the mSUGRA model. the SUSY contributions can
provide an additional $\sim 30\%$ enhancements, which play an
important rule in interpreting the $\eta' K$ puzzle. If we also
consider the effects of those dominant errors, the theoretical
predictions (in unit of $10^{-6}$ ) are \beq \label{eq:etap1}
Br(B^- \to K^- \eta')&=& \left\{ {\begin{array}{ll} 38.1\;
^{+9.4}_{-5.2}(\mu)\; ^{+6.6}_{-5.2} (F_0) \;
^{+13.2}_{-7.4}(\overline{m}_s)\; ^{+1.7}_{-1.2}(\gamma)
 & {\rm in \ \ SM},   \\
49.4\; ^{+11.1}_{-6.1}(\mu) \; ^{+9.6}_{-7.6} (F_0) \;
^{+15.7}_{-8.9}(\overline{m}_s)\; ^{+1.9}_{-1.3}(\gamma)
   & {\rm in \ \ mSUGRA},   \\
\end{array}} \right. \non
&=& \left\{ {\begin{array}{ll}
38.1 \; ^{+17.6}_{-10.5}  & {\rm in \ \ SM},   \\
49.4\; ^{+21.6}_{-13.3}   & {\rm in \ \ mSUGRA},   \\
\end{array}} \right. \\
Br(B^0 \to K^0 \eta')&=& \left\{ {\begin{array}{ll}
35.4\; ^{+8.9}_{-5.0}(\mu)\; ^{+6.4}_{-5.0} (F_0) \;
^{+11.9}_{-6.7}(\overline{m}_s)\; \pm 0.3 (\gamma)
 & {\rm in \ \ SM},   \\
45.7\; ^{+10.5}_{-5.9}(\mu)\; ^{+9.1}_{-7.3} (F_0) \;
^{+14.1}_{-8.0}(\overline{m}_s)\; \pm 0.3(\gamma)
   & {\rm in \ \ mSUGRA},   \\
\end{array}} \right. \non
&=& \left\{ {\begin{array}{ll}
35.4\; ^{+16.2}_{-9.7} & {\rm in \ \ SM},   \\
45.7\; ^{+19.8}_{-12.3}  & {\rm in \ \ mSUGRA},   \\
\end{array}} \right.
\label{eq:etap0}
\eeq
where the individual errors are added in quadrature.
The relation between $F_0^{B\to \eta^{(')}}$ and $F_0^{B \to \pi}$
have been defined in Ref.~\cite{tpb98}.
It is evident that the theoretical predictions in the mSUGRA model
are consistent with the data within one standard deviation.

\begin{figure}[thb]
\centerline{\mbox{ \epsfxsize=8cm\epsffile{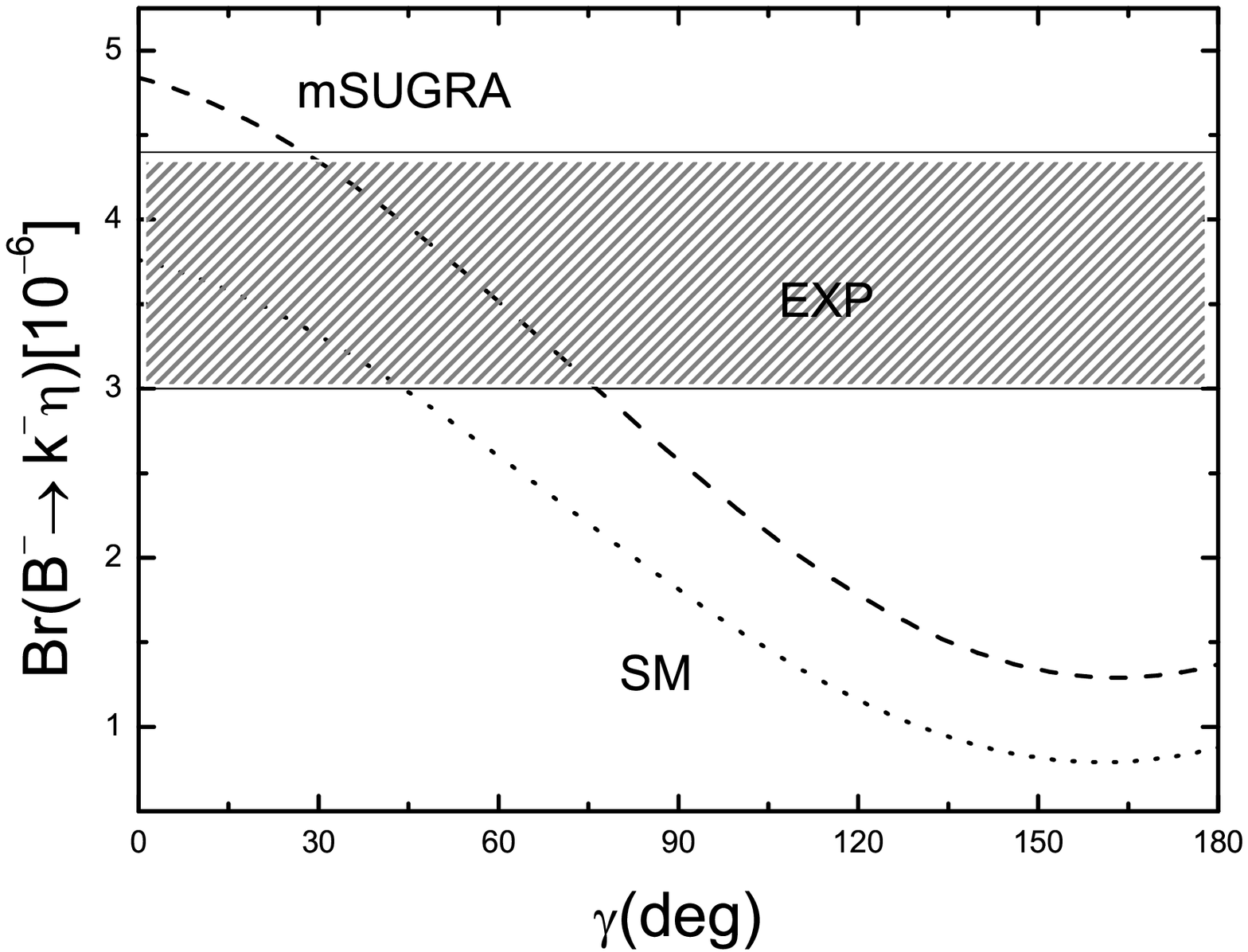}
\epsfxsize=8cm\epsffile{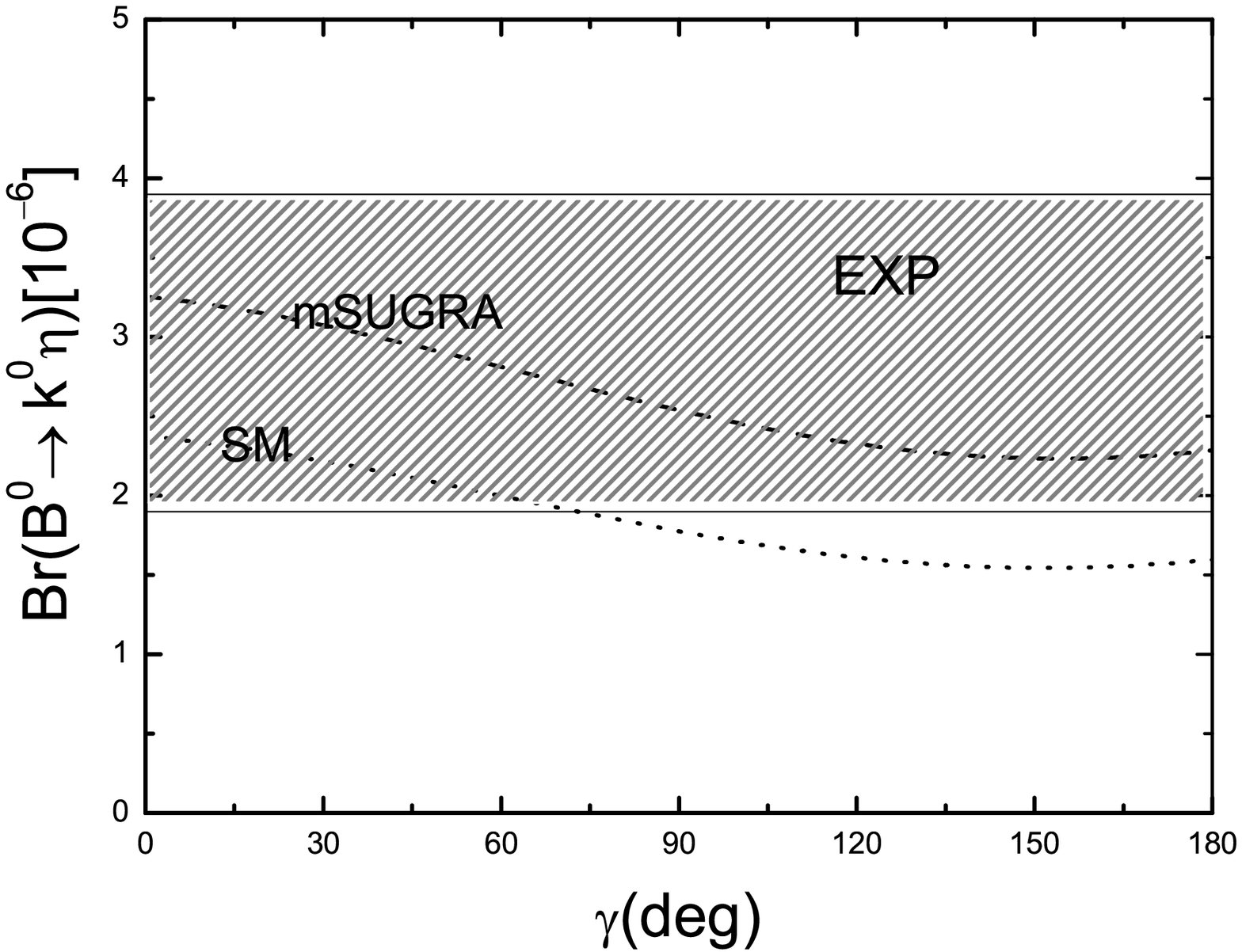}}} \vspace{0.1cm}
\centerline{\mbox{ \epsfxsize=8cm\epsffile{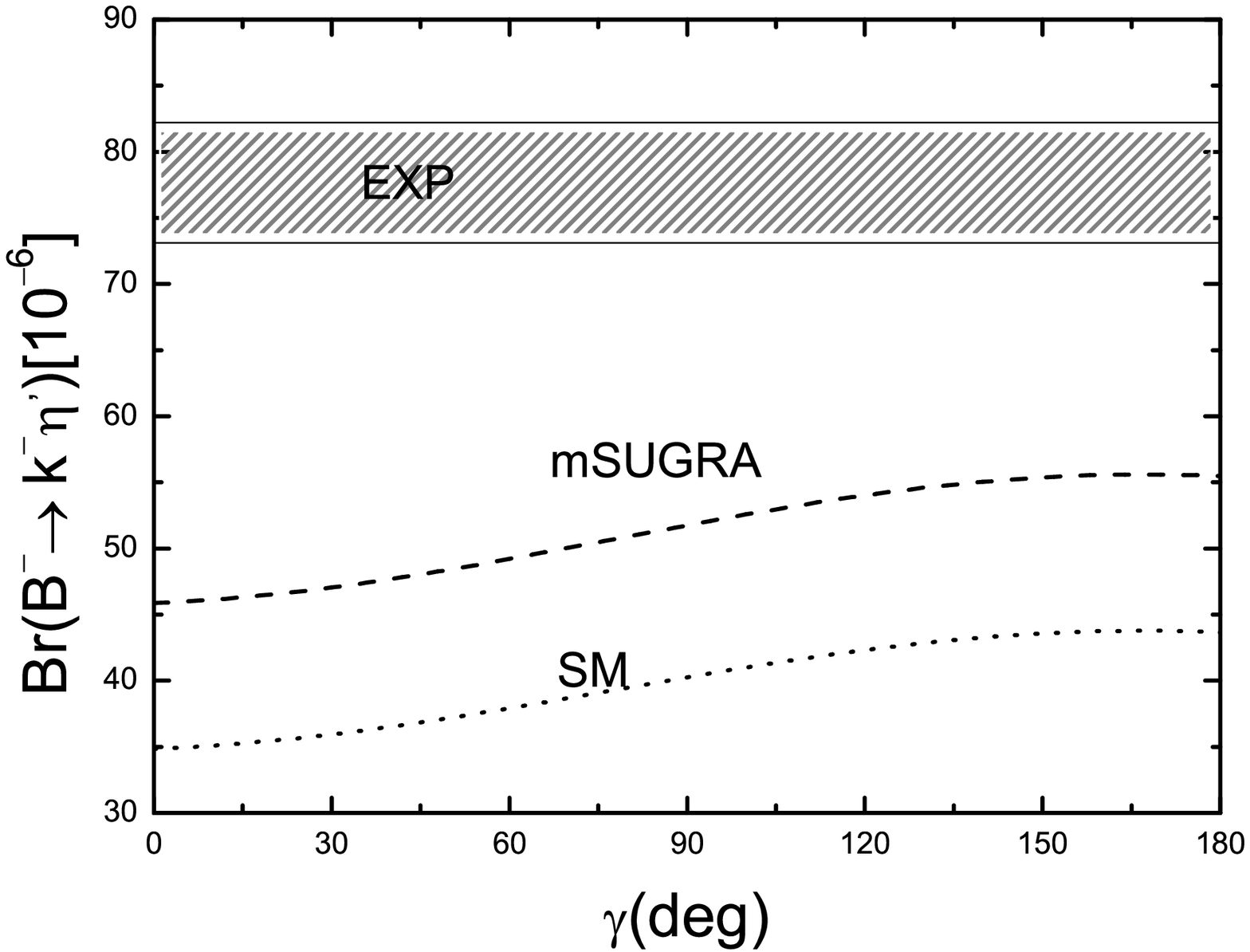}
\epsfxsize=8cm\epsffile{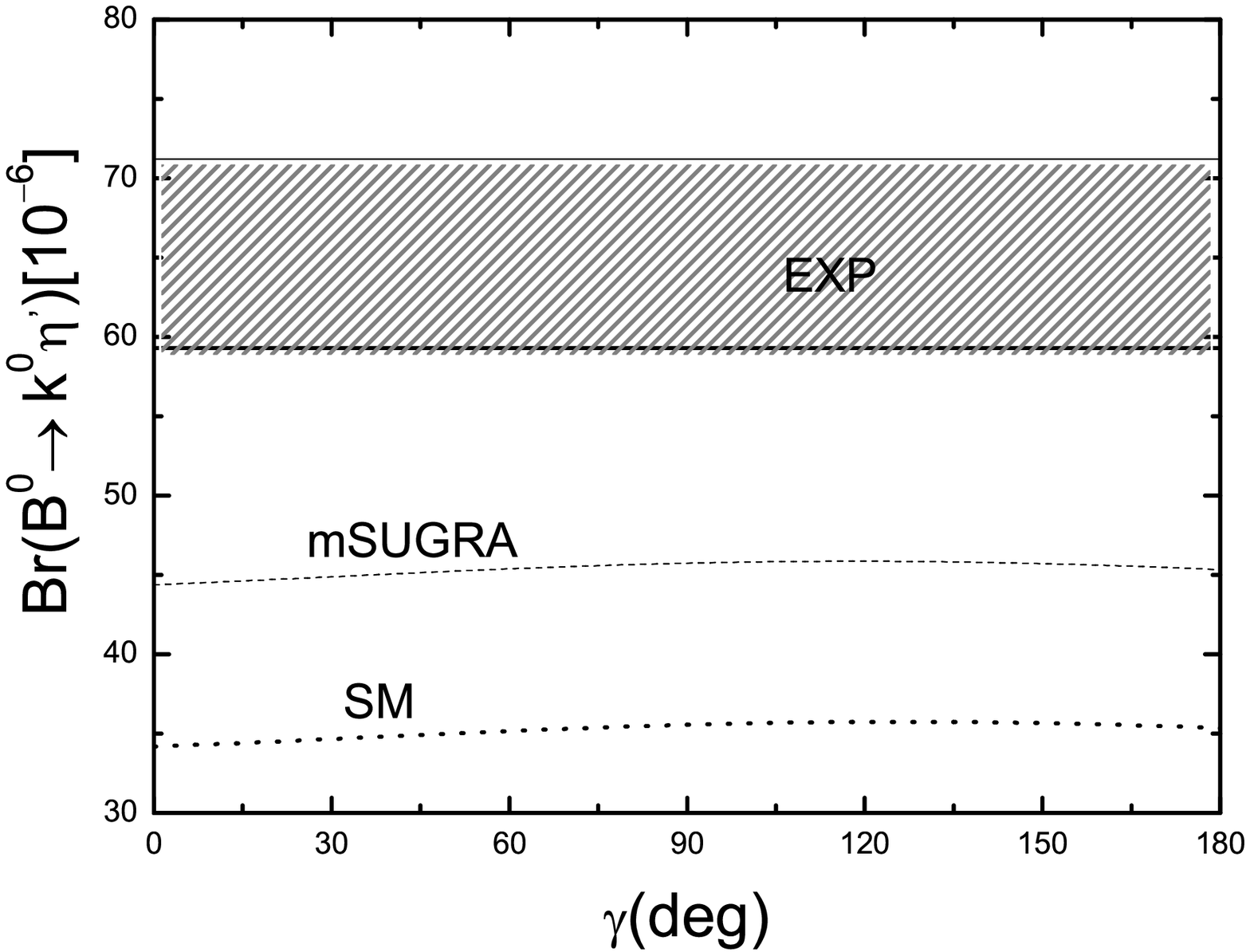}}} \vspace{0.1cm} \caption{ The
$\gamma$ dependence of the branching ratios $Br(B \to K
\eta^{(')})$ in the SM and  minimal SUGRA model. The dots and
solid curves show the central values of the SM and mSUGRA
predictions. The horizontal gray bands show the measured value as
given in Table III.}. \label{fig:keta}
\end{figure}

For $B^- \to K^- \eta$ and $B^0 \to K^0 \eta$ decays,
the annihilation contributions are less than $2\%$, while the SUSY
enhancements are about $30\%$. The numerical results (in unit of
$10^{-6}$) are
\beq
\label{eq:kpeta} Br(B^- \to K^- \eta)&=&
\left\{ {\begin{array}{ll} 2.7\;  ^{+0.0}_{-0.1}(\mu)\;
^{+1.0}_{-0.9} (F_0) \; ^{+1.1}_{-0.6} (\overline{m}_s)\;
^{+0.4}_{-0.6}(\gamma)
& {\rm in \ \ SM,}   \\
3.6\; ^{+0.0}_{-0.2}(\mu)\; ^{+1.3}_{-1.1} (F_0)
\; ^{+1.4}_{-0.8} (\overline{m}_s)\; ^{+0.5}_{-0.7}(\gamma)
  & {\rm in \ \ mSUGRA,}   \\
\end{array}} \right. \\
\label{eq:k0eta}
Br(B^0 \to K^0 \eta)&=& \left\{
{\begin{array}{ll} 2.0\;  ^{+0.0}_{-0.1}(\mu)\; ^{+0.7}_{-0.6}
(F_0) \; ^{+1.0}_{-0.6} (\overline{m}_s)\; ^{+0.1}_{-0.2}(\gamma)
& {\rm in \ \ SM,}   \\
2.8\; ^{+0.0}_{-0.1}(\mu)\; ^{+1.0}_{-0.8} (F_0)
\; ^{+1.2}_{-0.7} (\overline{m}_s)\; \pm 0.2 (\gamma)
  & {\rm in \ \ mSUGRA,}   \\
\end{array}} \right.
\eeq
The theoretical predictions in both the SM and mSUGURA model are
all consistent with the data within one standard deviation.
But the consistency between the theoretical predictions and the data
is clearly improved by the inclusion of the SUSY contribution.

In Fig.~\ref{fig:keta}, we show the $\gamma$ dependence of the
branching ratios for four  $B \to K\eta^{(')}$ decays. The dots
and dashed curves correspond to the central values of the
theoretical prediction in the SM and mSUGRA model, respectively.
The horizontal slashed bands show the data as given in Table
\ref{tab:exp}.

\subsection{$B \to \pi \eta^{(')}$ and $B \to \eta \eta^{(')}$}

These seven decay modes are tree-dominated decay processes, and
new physics enhancements due to the SUSY contributions are less
than $10\%$. For the measured $B^- \to \pi^- \eta$ decay, the
theoretical predictions in the SM and mSUGRA model are
\beq
\label{eq:pieta}
Br(B^- \to \pi^- \eta)&=& \left\{
{\begin{array}{ll} 4.4\; + 0.1(\mu)\; ^{+1.4}_{-1.2} (F_0) \;
^{+0.3}_{-0.2} (\overline{m}_s)\; ^{+0.9}_{-1.3}(\gamma)
& {\rm in \ \ SM}   \\
4.7\; + 0.1 (\mu)\; ^{+1.5}_{-1.3} (F_0)
\; ^{+0.3}_{-0.2} (\overline{m}_s)\; ^{+1.0}_{-1.5}(\gamma)
  & {\rm in \ \ mSUGRA}   \\
\end{array}} \right.
\eeq and consistent with the experimental measurements within one
standard deviation. For  $B^- \to \pi^- \eta'$ decay, the
theoretical prediction for its branching ratio is similar with
that for $B^- \to \pi^- \eta$ decay, and may be observed soon by
the B-factory experiments.

For $B^0 \to \pi^0 \eta^{(')}$ and $B^0 \to \eta^{(')}
\eta^{(')}$ decays, the theoretical predictions in the SM and
mSUGRA model are of order $10^{-7}$ and smaller than the
experimental upper limits.

For B meson decays involving an $\eta$ or $\eta^{'}$ as at least
one of the two final states, some specific contributions such as
the color singlet contribution have been discussed in
\cite{nb0325}. These contributions will be in favor of accounting
for the experimental data. However, large uncertainties go with
them. In our calculations, such contributions are not taken into
account.

\subsection{$B \to K K $ decays}

For the three $B \to K K$ decay modes, only experimental upper
limits are available now. The $B^0\to K^+K^{-}$ decay
receives only the weak annihilation contribution. It's branching
ratio is strongly suppressed in the QCD factorization approach. In
Ref.\cite{ch01},  the authors calculate this decay mode by
employing the PQCD approach and also found  a small branching
ratio. From this decay mode, we can obtain useful information
about the long-distance final state interaction and soft
annihilations when the precise experimental measurement becomes
available in the future.

For $B^0 \to K^0 \overline{K}^0$ and $B^\pm \to K^\pm K^0$
decays, they are penguin-dominant and the SUSY contributions can
provide $\sim 20\%$ enhancements to their branching ratios, and
still within the experimental upper limits.

\subsection{Uncertainties of theoretical predictions}

From the numerical results as given in Table \ref{tab:smmssm} and in
Eqs.(\ref{eq:pipi1}-\ref{eq:pieta}), one can see that
the theoretical predictions still have large uncertainties.

For most $B \to PP$ decays,  the dominant error comes from the
uncertainties of the corresponding form factors, since the
branching ratios are generally proportional to the square of the
related form factors. The measured $Br(B^0 \to \pi^+
\pi^-)=(4.6 \pm 0.4)\times 10^{-6}$ prefers a smaller $F_0^{B\to
\pi}(0)$, but the large decay rates for $B^0 \to \pi^0
\pi^0$ needs a large $F_0^{B\to \pi}(0)$. The measured large
branching ratios for $B \to K \pi$ and $K \eta'$ decays also favor
large $F_0^{B\to K}(0)$ and $F_0^{B\to \eta'}(0)$. Further
reduction of the uncertainties of the form factors is essential
for us to find the signal of new physics from the $B \to PP$
decays.

The large uncertainty of the light quark masses is also a major source of the
theoretical errors. For $B^-\to K^-\eta^{'}$, for example,
a $44\%$ enhancement can be obtained by varying  $\overline{m}_s(2GeV)$
from $105$ MeV to $80 $ MeV.

Thirdly, the CKM angle $\gamma$ has a wide scope and can bring large
uncertainties to the theoretical predictions for some $B \to PP$ decays
in both the SM and the mSUGRA model. Of course, one can also constrain
the angle $\gamma$ from the experimental measurements
of $B \to K \pi$ decays \cite{xz02}.

The $\gamma$ dependence of the branching ratios for those measured
$B \to PP$ decays are illustrated in Figs.~\ref{fig:pipi}-\ref{fig:keta}.
In these figures, the dots and dashed line shows  the SM  and the mSUGRA
predictions, respectively. The theoretical uncertainties are not explicitly
shown here.

From Figs.~\ref{fig:pipi}-\ref{fig:keta}, one  can see that some
decay modes ($B\to\pi^+\pi^-, \pi^0 K^-, \pi^+ K^-$, etc.) are
sensitive to the angle $\gamma$ in both SM and the minimal SUGRA
model, while other decays such as $B \to \pi^- K^0, \pi^0 K^0$ and
$K \eta'$ have a weak dependence on the angle $\gamma$. By
analyzing the expressions of the decay amplitudes, we find that if
the term proportional to $V_{ub}$ is dominant over other terms in
the total decay amplitude of a given decay, the branching ratio of
this decay will has a strong dependence on the angle $\gamma$.

Fourthly,  the endpoint divergence of $X_{H}$ in the hard spectator
scattering  can produce large uncertainty to the theoretical
calculations. But it is generally  not important
for those tree- or penguin-dominated decay processes because of
the strong suppression of the $\alpha_s$ and $N_c$.

In the QCD factorization approach, the annihilation contributions
can not be calculated reliably, but estimated with large uncertainty.
For $B \to PP$ decays, the annihilation contribution may be strongly
power suppressed as discussed by Ali et al.\cite{ali9804}.
Of course, such assumption has given rise to some controversy.

Finally, when considering the branching ratios in the minimal
SUGRA model, different numerical results can be obtained by
varying the SUSY parameters ($m_0,m_{1/2},\tan{\beta},A_0, Sign(\mu)$)
around the given values as listed in Table \ref{tab:mssm}.
In this paper, we considered two typical sets of SUSY parameters
which are still allowed by the data of $B \to X_s \gamma$ and
other measurements. In case A the SUSY contribution is small
and can hardly change the SM predictions.
On the contrary, in case B the SUSY contribution is significant in size
and provides favorable enhancements to the branching ratios of the
penguin dominant decay modes.

\section{Summary}

In this paper, we calculated the SUSY contributions to the branching
ratios of $B\to PP$ decays  in the framework of the mSUGRA model
by employing the QCD factorization approach.

In Sec.~\ref{sec:msugra}, a brief review about the mSUGRA model
was given. In Sec.~\ref{sec:tbtf}, we evaluated analytically the
new penguin diagrams induced by new particles ( gluinos,
charged-Higgs bosons, charginos and neutalinos), and obtained the
analytical expressions of the SUSY contributions to the Wilson
coefficients. The calculation of $B \to PP$ decays in the QCD
factorization approach is also discussed in this section. For the
mSUGRA model with the mixing matrix as given in Eq.(\ref{eq:ud}),
we found that (a) the SUSY corrections to the Wilson coefficients
$C_k$ ($k=3-6$) are very small and can be neglected safely; (b)
the leading order SUSY contributions  to the Wilson coefficients
$C_{7\gamma}(M_W)$  and $C_{8g}(M_W)$ can be rather large, and
even change the sign of the corresponding coefficients in the SM.

In Sec.~\ref{sec:nc}, we calculated the branching ratios
for twenty one $B  \to PP$ decays in the SM and the mSUGRA model,
and made phenomenological analysis for some well measured decay modes.
From the numerical results, we find following general features about the new
physics effects on the exclusive charmless hadronic $B \to PP$ decays
studied in this paper:

\begin{enumerate}

\item
For those tree-dominated decays, such as $B \to \pi\pi$, the possible
SUSY contributions in mSUGRA model are very small and can be neglected
safely.

\item
For those  penguin-dominated decay modes, the SUSY contributions to their
branching ratios can be significant, around $30-50\%$.

\item
For the four $B \to K \pi$ decays, the SUSY contributions to the branching ratios
play an important rule to improve the consistency of the theoretical predictions
with the data.

\item
For $B \to K \eta'$ decays, the theoretical predictions for branching ratios
become consistent with the measured values within one standard deviation
after the inclusion of the large SUSY contributions in the mSUGRA model.
This is a possible interpretation for the so-called $K \eta'$ puzzle.

\item
The theoretical predictions in both the SM and mSUGRA model still have large
theoretical uncertainties. The dominant errors are induced by the
uncertainties of
the form factors $F_0^{B \to P}$, strange quark mass $\overline{m}_s$,
the low-energy scale $\mu \sim m_b$ and the CKM angle $\gamma$.

\end{enumerate}

\begin{acknowledgments}

We are very grateful to Cai-dian L\"u, Li-bo Guo, Chao-shang Huang,
Xiao-hong Wu, Li-gang Jin, and Ying Li for helpful discussions.
This work is partially supported  by the National
Natural Science Foundation of China under Grant No.10275035,
and by the Research Foundation of Nanjing Normal University under
Grant No.~214080A916.

\end{acknowledgments}


\begin{appendix}

\section{One-Loop function and the coupling constants} \label{app:fun}

In this Appendix, the explicit expressions of
$f_i(x)$ functions and the coupling constants appeared
in Eq.(\ref{eq:f1lh})-(\ref{eq:f2r}) are presented.
For more details, one can see Ref.\cite{pm96} and references therein.
\beq
f_{1}(x)&=&\frac{1}{12(x-1)^{4}}\left (x^3-6x^2+3x+2+6x \ln x \right ), \\
f_{2}(x)&=&\frac{1}{12(x-1)^{4}}\left( 2x^3+3x^2-6x+1-6x^2 \ln x\right ),\\
f_{3}(x)&=&\frac{1}{2(x-1)^{3}}\left ( x^2-4x+3+2 \ln x \right ),          \\
f_{4}(x)&=&\frac{1}{2(x-1)^{3}}\left(x^2-1-2x \ln x\right ),              \\
f_{5}(x)&=&\frac{1}{36(x-1)^{4}}\left (7x^3-36x^2+45x-16-12 \ln x+18x\ln x \right ),
\\
f_{6}(x)&=&\frac{1}{54(x-1)^{4}}\left (
37-171x+207x^2-73x^3+3\ln x-81x^2\ln x+54x^3\ln x \right ), \\
f_{7}(x)&=&\frac{1}{18(x-1)^{4}}\left (-11x^3+18x^2-9x+2+6x^3\ln x \right ).
\eeq
For the coupling constants $\Gamma^{d}_{G(LR)}, \Gamma^{d}_{C(LR)}, \Gamma^{d}_{N(LR)}$, we have
\beq
\left (\Gamma^{d}_{GL}\right )_{I}^{j}&=&\left (\Gamma^{D} \right )^{j}_{I},\\
\left (\Gamma^{d}_{GR}\right )_{I}^{j}&=&-\left (\Gamma^{D}\right )^{j+3}_{I},\\
\left (\Gamma^{d}_{CL}\right )_{I}^{\alpha j}&=&
\left [V^{*\alpha}_{1}\left (\Gamma^{U}\right )^{k}_{I}-
V^{*\alpha}_{2}\left (\Gamma^{U} \right )^{k+3}_{I}\frac{m_{k}^{u}}{\sqrt{2}m_{W}
\sin{\beta}}\right ]K_{kj},
\\
\left (\Gamma^{d}_{CR} \right )_{I}^{\alpha j}&=&-U^{\alpha}_{2}(\Gamma^{U})^{k}_{I}
\frac{m_{k}^{d}}{\sqrt{2}m_{W}\cos{\beta}}K_{kj},
\\
\left (\Gamma^{d}_{NL} \right )_{I}^{\alpha j}&=&
\frac{1}{\sqrt{2}}\left [\left (-N^{*\alpha}_{2}+\frac{1}{3}
\tan{\theta_W}N^{*\alpha}_{1}\right )\left (\Gamma^{D}\right )^{j}_{I} +
N^{*\alpha}_{3}\left (\Gamma^{D}\right )^{j+3}_{I}\frac{m^{d}_{j}}{m_{W}
\cos{\beta}}\right ],\\
\left (\Gamma^{d}_{NR}\right )_{I}^{\alpha
j}&=&\frac{1}{\sqrt{2}}\left
[\frac{2}{3}\tan{\theta_W}N^{\alpha}_{1} \left (\Gamma^{D}\right
)^{j+3}_{I}+ N^{\alpha}_{3}\left (\Gamma^{D}\right
)^{j}_{I}\frac{m^{d}_{j}}{m_{W}\cos{\beta}}\right ],
\eeq
where the K is the CKM matrix.

\end{appendix}


\end{document}